%% file: ms.tex
\pdfoutput=1
\newif\ifDraft\Drafttrue

\documentclass[letterpaper,twocolumn,10pt]{article}
\usepackage{usenix,epsfig,endnotes}
\usepackage{tugraz_defaults}
\usepackage{csquotes}
\usepackage{enumitem}
\usepackage{fontawesome}

\newcommand{\RNum}[1]{\uppercase\expandafter{\romannumeral #1\relax}}

\renewcommand{\paragraph}[1]{\vspace{0.0cm}\noindent\textbf{#1}\ }
\renewcommand{\subparagraph}[1]{\vspace{0.0cm}\noindent\textbf{#1}\ }

\begin{document}


\date{}

\title{A Systematic Evaluation of Transient Execution Attacks and Defenses}

\author{{\rm Claudio Canella}$^1$, {\rm Jo Van Bulck}$^2$, {\rm Michael Schwarz}$^1$, {\rm Moritz Lipp}$^1$, \\ {\rm Benjamin von Berg}$^1$, {\rm Philipp Ortner}$^1$, {\rm Frank Piessens}$^2$, {\rm Dmitry Evtyushkin}$^3$, {\rm Daniel Gruss}$^1$ \\
$^1$ Graz University of Technology, $^2$ imec-DistriNet, KU Leuven, $^3$ College of William and Mary }

\maketitle

\pagestyle{empty}

\newcommand{\Spectre}{Spectre\xspace}
\newcommand{\VariantOne}{Variant 1\xspace}
\newcommand{\VariantOneOne}{Variant 1.1\xspace}
\newcommand{\SpectrePHT}{Spectre-PHT\xspace}

\newcommand{\VariantTwo}{Variant 2\xspace}
\newcommand{\SpectreBTB}{Spectre-BTB\xspace}

\newcommand{\VariantThree}{Variant 3\xspace}

\newcommand{\VariantFour}{Variant 4\xspace}
\newcommand{\SpectreSTL}{Spectre-STL\xspace}

\newcommand{\SpectreRetToSpec}{ret2spec\xspace}
\newcommand{\SpectreRSB}{Spectre-RSB\xspace}

\newcommand{\NetSpectre}{NetSpectre\xspace}

\newcommand{\Meltdown}{Meltdown\xspace}
\newcommand{\MD}{\Meltdown}
\newcommand{\MeltdownA}{Variant 3a\xspace}
\newcommand{\Foreshadow}{Foreshadow\xspace}
\newcommand{\FS}{\Foreshadow}
\newcommand{\ForeshadowNG}{Foreshadow-NG\xspace}
\newcommand{\LazyFP}{Lazy FP\xspace}

\newcommand{\VariantOneTwo}{Variant 1.2\xspace}

\newcommand{\MeltdownDE}{Meltdown-DE\xspace}
\newcommand{\MeltdownAC}{Meltdown-AC\xspace}
\newcommand{\MeltdownBR}{Meltdown-BR\xspace}
\newcommand{\MeltdownUD}{Meltdown-UD\xspace}
\newcommand{\MeltdownGP}{Meltdown-GP\xspace}
\newcommand{\MeltdownNM}{Meltdown-NM\xspace}
\newcommand{\MeltdownSS}{Meltdown-SS\xspace}
\newcommand{\MeltdownSM}{Meltdown-SM\xspace}

\newcommand{\MeltdownUS}{Meltdown-US\xspace}
\newcommand{\MeltdownP}{Meltdown-P\xspace}
\newcommand{\MeltdownRW}{Meltdown-RW\xspace}
\newcommand{\MeltdownPK}{Meltdown-PK\xspace}
\newcommand{\MeltdownXD}{Meltdown-XD\xspace}
\newcommand{\MeltdownSMAP}{Meltdown-SMAP\xspace}
\newcommand{\MeltdownSMEP}{Meltdown-SMEP\xspace}

\newcommand{\PF}{\texttt{\#PF}\xspace}
\newcommand{\GP}{\texttt{\#GP}\xspace}
\newcommand{\NM}{\texttt{\#NM}\xspace}
\newcommand{\BR}{\texttt{\#BR}\xspace}
\newcommand{\DE}{\texttt{\#DE}\xspace}
\newcommand{\UD}{\texttt{\#UD}\xspace}
\newcommand{\SF}{\texttt{\#SS}\xspace}
\newcommand{\AC}{\texttt{\#AC}\xspace}

\newcommand{\inst}[1]{\texttt{#1}}
\newcommand{\rdclno}{{RDCL\_NO}\xspace}


\begin{abstract}
Research on \emph{transient execution} attacks including Spectre and Meltdown showed that exception or branch misprediction events might leave secret-dependent traces in the CPU's microarchitectural state.
This observation led to a proliferation of new Spectre and Meltdown attack variants and even more ad-hoc defenses (\eg microcode and software patches).
Both the industry and academia are now focusing on finding effective defenses for known issues.
However, we only have limited insight on residual attack surface and the completeness of the proposed defenses.

In this paper, we present a systematization of transient execution attacks.
Our systematization uncovers 6 (new) transient execution attacks that have been overlooked and not been investigated so far:
  2 new exploitable Meltdown effects: \MeltdownPK (Protection Key Bypass) on Intel, and Meltdown-BND (Bounds Check Bypass) on Intel and AMD;
and 4 new Spectre mistraining strategies.
We evaluate the attacks in our classification tree through proof-of-concept implementations on 3 major CPU vendors (Intel, AMD, ARM).
Our systematization yields a more complete picture of the attack surface and allows for a more systematic evaluation of defenses.
Through this systematic evaluation, we discover that most defenses, including deployed ones, cannot fully mitigate all attack variants.
\end{abstract}


\section{Introduction}
CPU performance over the last decades was continuously improved by shrinking processing technology and increasing clock frequencies, but physical limitations are already hindering this approach.
To still increase the performance, vendors shifted the focus to increasing the number of cores and optimizing the instruction pipeline.
Modern CPU pipelines are massively parallelized allowing hardware logic in prior pipeline stages to perform operations for subsequent instructions ahead of time or even out-of-order.
Intuitively, pipelines may stall when operations have a dependency on a previous instruction which has not been executed (and retired) yet.
Hence, to keep the pipeline full at all times, it is essential to predict the control flow, data dependencies, and possibly even the actual data.
Modern CPUs, therefore, rely on intricate microarchitectural optimizations to predict and sometimes even re-order the instruction stream.
Crucially, however, as these predictions may turn out to be wrong, pipeline flushes may be necessary, and instruction results should always be committed according to the intended in-order instruction stream.
Pipeline flushes may occur even without prediction mechanisms, as on modern CPUs virtually any instruction can raise a fault (\eg page fault or general protection fault), requiring a roll-back of all operations following the faulting instruction.
With prediction mechanisms, there are more situations when partial pipeline flushes are necessary, namely on every misprediction.
The pipeline flush discards any architectural effects of pending instructions, ensuring functional correctness.
Hence, the instructions are executed \emph{transiently} (first they are, and then they vanish), \ie we call this \emph{transient execution}~\cite{Lipp2018meltdown,Kocher2019spectre,Vanbulck2018}.

While the architectural effects and results of transient instructions are discarded, microarchitectural side effects remain beyond the transient execution.
This is the foundation of \Spectre~\cite{Kocher2019spectre}, \Meltdown~\cite{Lipp2018meltdown}, and Foreshadow~\cite{Vanbulck2018}.
These attacks exploit transient execution to encode secrets through microarchitectural side effects (\eg cache state) that can later be recovered by an attacker at the architectural level.
The field of transient execution attacks emerged suddenly and proliferated, leading to a situation where people are not aware of all variants and their implications.
This is apparent from the confusing naming scheme that already led to an arguably wrong classification of at least one attack~\cite{Kiriansky2018speculative}.
Even more important, this confusion leads to misconceptions and wrong assumptions for defenses.
Many defenses focus exclusively on hindering exploitation of a specific covert channel, instead of addressing the microarchitectural root cause of the leakage~\cite{Kiriansky2018dawg, khasawneh2018safespec, Yan2018InvisiSpec, Kocher2019spectre}.
Other defenses rely on recent CPU features that have not yet been evaluated from a transient security perspective~\cite{VahldiekOberwagner2018}.
We also debunk implicit assumptions including that AMD or the latest Intel CPUs are completely immune to \Meltdown-type effects, or that serializing instructions mitigate \Spectre\VariantOne on any CPU.

\begin{figure}
    \centering
    \resizebox{\hsize}{!}{
     \input{images/classification_small.tikz}
    }
    \caption{Transient execution attack classification tree with
             demonstrated attacks (red, bold), negative results (green, dashed), some first explored in this work ($\medblackstar$ / $\medwhitestar$).}
    \label{fig:tree}
    \vspace{-0.2cm}
\end{figure}
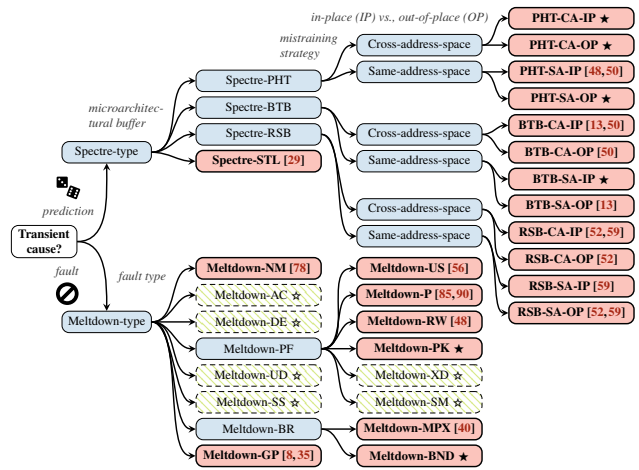

In this paper, we present a systematization of transient execution attacks, \ie Spectre, Meltdown, Foreshadow, and related attacks.
Using our decision tree, transient execution attacks are accurately classified through an unambiguous naming scheme (\cf \cref{fig:tree}).
The hierarchical and extensible nature of our taxonomy allows to easily identify residual attack surface, leading to 6 previously overlooked transient execution attacks (Spectre and Meltdown variants) first described in this work.
Two of the attacks are Meltdown-BND, exploiting a Meltdown-type effect on the x86 \inst{bound} instruction on Intel and AMD, and \MeltdownPK, exploiting a Meltdown-type effect on memory protection keys on Intel.
The other 4 attacks are previously overlooked mistraining strategies for \SpectrePHT and \SpectreBTB attacks.
We demonstrate the attacks in our classification tree through practical proofs-of-concept with vulnerable code patterns evaluated on CPUs of Intel, ARM, and AMD.

Next, we provide a systematization of the state-of-the-art defenses.
Based on this, we systematically evaluate defenses with practical experiments and theoretical arguments to show which work and which do not or cannot suffice.
This systematic evaluation revealed that we can still mount transient execution attacks that are supposed to be mitigated by rolled out patches.
Finally, we discuss how defenses can be designed to mitigate entire types of transient execution attacks.

\paragraph{Contributions.}
The contributions of this work are:
\begin{enumerate}[nolistsep,align=left, leftmargin=13pt, labelwidth=0pt, itemindent=!]
    \item We systematize Spectre- and Meltdown-type attacks, advancing attack surface understanding, highlighting misclassifications, and revealing new attacks.
    \item We provide a clear distinction between Meltdown/Spectre, required for designing effective countermeasures.
    \item We categorize defenses and show that most, including deployed ones, cannot fully mitigate all attack variants.
    \item We describe new branch mistraining strategies, highlighting the difficulty of eradicating Spectre-type attacks.
\end{enumerate}
We responsibly disclosed the work to Intel, ARM, and AMD.

\paragraph{Experimental Setup.}
Unless noted otherwise, the experimental results reported were performed on recent Intel Skylake i5-6200U, Coffee Lake i7-8700K, and Whiskey Lake i7-8565U CPUs.
Our AMD test machines were a Ryzen 1950X and a Ryzen Threadripper 1920X.
For experiments on ARM, an NVIDIA Jetson TX1 has been used.

\paragraph{Outline.}
\cref{sec:background} provides background.
We systematize \Spectre in \cref{sec:spectre} and \Meltdown in \cref{sec:meltdown}.
We analyze and classify gadgets in \cref{sec:prevalence} and defenses in \cref{sec:defenses}.
We discuss future work and conclude in \cref{sec:future-work-conclusion}.


\section{Transient Execution}
\label{sec:background}
\paragraph{Instruction Set Architecture and Microarchitecture.}
The instruction set architecture (ISA) provides an interface between hardware and software.
It defines the instructions that a processor supports, the available registers, the addressing mode, and describes the execution model.
Examples of different ISAs are x86 and ARMv8.
The microarchitecture then describes how the ISA is implemented in a processor in the form of pipeline depth, interconnection of elements, execution units, cache, branch prediction.
The ISA and the microarchitecture are both stateful.
In the ISA, this state includes, for instance, data in registers or main memory after a successful computation.
Therefore, the architectural state can be observed by the developer.
The microarchitectural state includes, for instance, entries in the cache and the translation lookaside buffer (TLB), or the usage of the execution units.
Those microarchitectural elements are transparent to the programmer and can not be observed directly, only indirectly.

\paragraph{Out-of-Order Execution.}
On modern CPUs, individual instructions of a complex instruction set are first decoded and split-up into simpler micro-operations (\muops) that are then processed.
This design decision allows for superscalar optimizations and to extend or modify the implementation of specific instructions through so-called microcode updates.
Furthermore, to increase performance, CPU's usually implement a so-called \textit{out-of-order} design.
This allows the CPU to execute \muops not only in the sequential order provided by the instruction stream but to dispatch them in parallel, utilizing the CPU's execution units as much as possible and, thus, improving the overall performance.
If the required operands of a \muop are available, and its corresponding execution unit is not busy, the CPU starts its execution even if \muops earlier in the instruction stream have not finished yet.
As immediate results are only made visible at the architectural level when all previous \muops have finished, CPUs typically keep track of the status of \muops in a so-called \textit{Reorder Buffer} (ROB).
The CPU takes care to \emph{retire} \muops in-order, deciding to either discard their results or commit them to the architectural state.
For instance, exceptions and external interrupt requests are handled during retirement by flushing any outstanding \muop results from the ROB.
Therefore, the CPU may have executed so-called \textit{transient instructions}~\cite{Lipp2018meltdown}, whose results are never committed to the architectural state.

\paragraph{Speculative Execution.}
Software is mostly not linear but contains (conditional) branches or data dependencies between instructions.
In theory, the CPU would have to stall until a branch or dependencies are resolved before it can continue the execution.
As stalling decreases performance significantly, CPUs deploy various mechanisms to predict the outcome of a branch or a data dependency.
Thus, CPUs continue executing along the predicted path, buffering the results in the ROB until the correctness of the prediction is verified as its dependencies are resolved.
In the case of a correct prediction, the CPU can commit the pre-computed results from the reorder buffer, increasing the overall performance.
However, if the prediction was incorrect, the CPU needs to perform a roll-back to the last correct state by squashing all pre-computed transient instruction results from the ROB.

\paragraph{Cache Covert Channels.}
Modern CPUs use caches to hide memory latency.
However, these latency differences can be exploited in side-channels and covert channels~\cite{Kocher1996,Osvik2006,Yarom2014,Gruss2015Template,Maurice2017Hello}.
In particular, \FlushReload allows observations across cores at cache-line granularity, enabling attacks, \eg on cryptographic algorithms~\cite{Yarom2014,Irazoqui2014,Guelmezoglu2015}, user input~\cite{Gruss2015Template,Lipp2016,Schwarz2018KeyDrown}, and kernel addressing information~\cite{Gruss2016Prefetch}.
For \FlushReload, the attacker continuously flushes a shared memory address using the \clflush instruction and afterward reloads the data.
If the victim used the cache line, accessing it will be fast; otherwise, it will be slow.

Covert channels are a special use case of side-channel attacks, where the attacker controls both the sender and the receiver.
This allows an attacker to bypass many restrictions that exist at the architectural level to leak information.

\begin{figure}
\centering
    \includegraphics[width=\hsize]{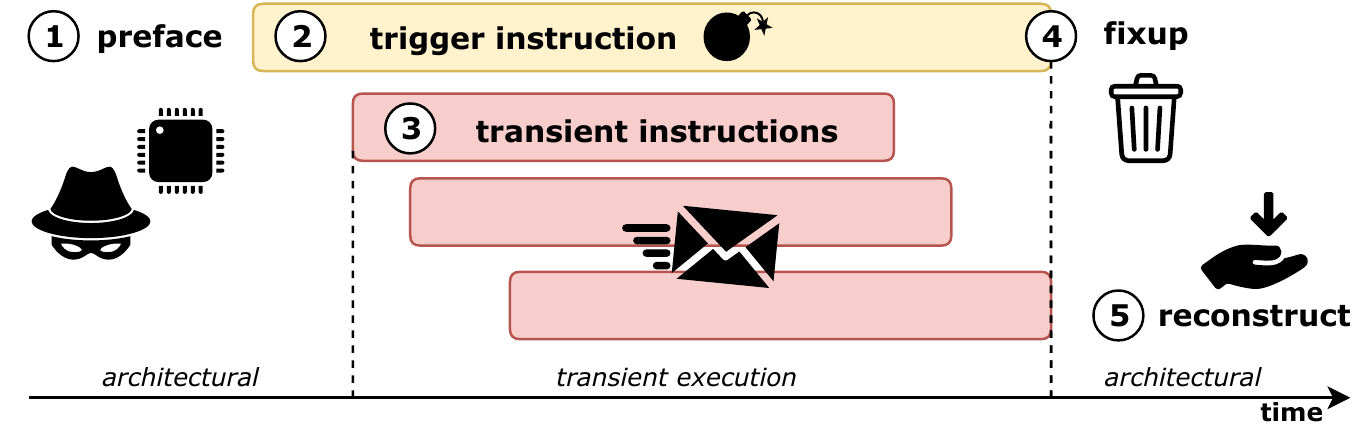}
    \caption{High-level overview of a transient execution attack in 5 phases:
                (1) prepare microarchitecture,
                (2) execute a {\em trigger instruction},
                (3) {\em transient instructions} encode unauthorized
                    data through a microarchitectural covert channel,
                (4) CPU retires trigger instruction and flushes
                    transient instructions,
                (5) reconstruct secret from microarchitectural state.
            }
    \label{fig:transient-flow}
    \vspace{-0.3cm}
\end{figure}

\paragraph{Transient Execution Attacks.}\label{subsec:tea}
Transient instructions reflect unauthorized computations out of the program's intended code and/or data paths.
For functional correctness, it is crucial that their results are never committed to the architectural state.
However, transient instructions may still leave traces in the CPU's microarchitectural state, which can subsequently be exploited to partially recover unauthorized results~\cite{Lipp2018meltdown,Kocher2019spectre,Vanbulck2018}.
This observation has led to a variety of transient execution attacks, which from a high-level always follow the same abstract flow, as shown in \Cref{fig:transient-flow}.

The attacker first brings the microarchitecture into the desired state, \eg by flushing and/or populating internal branch predictors or data caches.
Next is the execution of a so-called \emph{trigger instruction}.
This can be any instruction that causes subsequent operations to be eventually squashed, \eg due to an exception or a mispredicted branch or data dependency.
Before completion of the trigger instruction, the CPU proceeds with the execution of a \emph{transient instruction sequence}.
The attacker abuses the transient instructions to act as the sending end of a microarchitectural covert channel, \eg by loading a secret-dependent memory location into the CPU cache.
Ultimately, at the retirement of the trigger instruction, the CPU discovers the exception/misprediction and flushes the pipeline to discard any architectural effects of the transient instructions.
However, in the final phase of the attack, unauthorized transient computation results are recovered at the receiving end of the covert channel, \eg by timing memory accesses to deduce the secret-dependent loads from the transient instructions.

\paragraph{High-Level Classification: Spectre vs.\ Meltdown.}
Transient execution attacks have in common that they abuse transient instructions (which are never architecturally committed) to encode unauthorized data in the microarchitectural state.
With different instantiations of the abstract phases in \cref{fig:transient-flow}, a wide spectrum of transient execution attack variants emerges.
We deliberately based our classification on the root cause of the transient computation (phases 1, 2), abstracting away from the specific covert channel being used to transmit the unauthorized data (phases 3, 5).
This leads to a first important split in our classification tree (\cf \cref{fig:tree}).
Attacks of the first type, dubbed Spectre~\cite{Kocher2019spectre}, exploit transient execution following control or data flow misprediction.
Attacks of the second type, dubbed Meltdown~\cite{Lipp2018meltdown}, exploit transient execution following a faulting instruction.

Importantly, Spectre and Meltdown exploit fundamentally different CPU properties and hence require orthogonal defenses.
Where the former relies on dedicated control or data flow prediction machinery, the latter merely exploits that data from a faulting instruction is forwarded to instructions ahead in the pipeline.
Note that, while Meltdown-type attacks so far exploit out-of-order execution, even elementary in-order pipelines may allow for similar effects~\cite{Vanbulck2018nemesis}.
Essentially, the different root cause of the trigger instruction (Spectre-type misprediction vs.\ Meltdown-type fault) determines the nature of the subsequent unauthorized transient computations and hence the scope of the attack.

That is, in the case of Spectre, transient instructions can only compute on data which the application is also allowed to access architecturally.
Spectre thus transiently bypasses \emph{software-defined} security policies (\eg bounds checking, function call/return abstractions, memory stores) to leak secrets out of the program's intended code/data paths.
Hence, much like in a \enquote{confused deputy} scenario, successful Spectre attacks come down to steering a victim into transiently computing on memory locations the victim is authorized to access but the attacker not.
In practice, this implies that one or more phases of the transient execution attack flow in \cref{fig:transient-flow} should be realized through so-called \emph{code gadgets} executing within the victim application.
We propose a novel taxonomy of gadgets based on these phases in \cref{sec:prevalence}.

For Meltdown-type attacks, on the other hand, transient execution allows to completely ``melt down'' architectural isolation barriers by computing on unauthorized results
of faulting instructions.
Meltdown thus transiently bypasses \emph{hardware-enforced} security policies to leak data that should always remain architecturally inaccessible for the application.
Where Spectre-type leakage remains largely an unintended side-effect of important speculative performance optimizations, Meltdown reflects a failure of the CPU to respect hardware-level protection boundaries for transient instructions.
That is, the mere continuation of the transient execution after a fault itself is required, but not sufficient for a successful \Meltdown attack.
As further explored in \cref{sec:defenses}, this has profound consequences for defenses.
Overall, mitigating Spectre requires careful hardware-software co-design, whereas merely replacing the data of a faulting instruction with a dummy value suffices to block \Meltdown-type leakage in silicon, \eg as it is done in AMD processors, or with the Rogue Data Cache Load resistance (\rdclno{}) feature advertised in recent Intel CPUs from Whiskey Lake onwards~\cite{IntelMitigations}.

\begin{table}[t]
	\setlength{\aboverulesep}{0pt}
	\setlength{\belowrulesep}{0pt}
    \caption{Spectre-type attacks and the microarchitectural element they
             exploit (\circletfill), partially target (\circletfillhl), or not affect (\circlet).}\label{tab:spectre_attacks_me_target}
\begin{center}
\vspace{-0.375cm}
\adjustbox{max width=\columnwidth}{\small{
		\setlength\tabcolsep{1.5pt}
		\begin{tabular}{r llllll}
			\diagbox{\textbf{Attack}}{\textbf{Element}} &\,\, & \rotatebox{90}{\hspace{-0.66em}BTB} \hspace{-20em} & \rotatebox{90}{\hspace{-0.66em}BHB} \hspace{-20em} & \rotatebox{90}{\hspace{-0.66em}PHT} \hspace{-20em} & \rotatebox{90}{\hspace{-0.66em}RSB} \hspace{-20em} & \rotatebox{90}{\hspace{-0.66em}STL} \hspace{-20em}\\
			\toprule \vspace{-0.3cm}\\
			\SpectrePHT (\VariantOne)~\cite{Kocher2019spectre} &		 		& \cmarkempty & \circletfillhl 	& \circletfill 	& \cmarkempty 	& \cmarkempty \\
			\SpectrePHT (\VariantOneOne)~\cite{Kiriansky2018speculative} &	& \cmarkempty & \circletfillhl & \circletfill 	& \cmarkempty 	& \cmarkempty \\
			\SpectreBTB (\VariantTwo)~\cite{Kocher2019spectre} &		 		& \circletfill & \circletfillhl	& \cmarkempty 	& \cmarkempty 	& \cmarkempty \\
			\SpectreRSB(\SpectreRetToSpec)~\cite{Koruyeh2018spectre5,Maisuradze2018spectre5} &	& \circletfillhl & \cmarkempty	& \cmarkempty 	& \circletfill 	& \cmarkempty \\
			\SpectreSTL (\VariantFour)~\cite{Horn2018spectre4} &		 		& \cmarkempty & \cmarkempty	& \cmarkempty 	& \cmarkempty 	& \circletfill \\
			\bottomrule
		\end{tabular}
	}}
	\end{center}
\vspace{-0.3cm}
    {\footnotesize
    Glossary: Branch Target Buffer (BTB), Branch History Buffer (BHB), Pattern History Table (PHT), Return Stack Buffer (RSB), Store To Load (STL).
    }
    \vspace{-0.4cm}
\end{table}

\section{\Spectre-type Attacks}
\label{sec:spectre}
In this section, we provide an overview of \Spectre-type attacks (\cf \cref{fig:tree}).
Given the versatility of Spectre variants in a variety of adversary models, we propose a novel two-level taxonomy based on the preparatory phases of the abstract transient execution attack flow in \cref{fig:transient-flow}.
First, we distinguish the different microarchitectural buffers that can trigger a prediction (phase 2), and second, the mistraining strategies that can be used to steer the prediction (phase 1).

\paragraph{Systematization of Spectre Variants.}
To predict the outcome of various types of branches and data dependencies, modern CPUs accumulate an extensive microarchitectural state across various internal buffers and components~\cite{Fog2016}.
\Cref{tab:spectre_attacks_me_target} overviews \Spectre-type attacks and the corresponding microarchitectural elements they exploit.
As the first level of our classification tree, we categorize Spectre attacks based on the microarchitectural root cause that triggers the misprediction leading to the transient execution:
\begin{compactitem}
  \item \SpectrePHT~\cite{Kocher2019spectre,Kiriansky2018speculative} exploits the \textit{Pattern History Table} (PHT) that predicts the outcome of conditional branches.
  \item \SpectreBTB~\cite{Kocher2019spectre} exploits the \textit{Branch Target Buffer} (BTB) for predicting branch destination addresses.
  \item \SpectreRSB~\cite{Maisuradze2018spectre5,Koruyeh2018spectre5} primarily exploits the \textit{Return Stack Buffer} (RSB) for predicting return addresses.
  \item \SpectreSTL~\cite{Horn2018spectre4} exploits memory disambiguation for predicting \textit{Store To Load} (STL) data dependencies.
\end{compactitem}
Note that \NetSpectre~\cite{Schwarz2018netspectre}, SGXSpectre~\cite{OKeeffe18sgxspectre}, and SGXPectre~\cite{Chen2018SGXpectre} focus on applying one of the above \Spectre variants in a specific exploitation scenario.
Hence, we do not consider them separate variants in our classification.

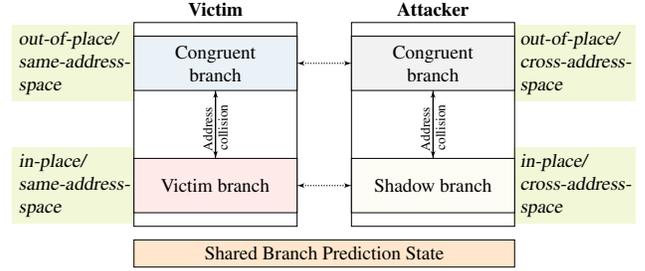
\begin{figure}
\adjustbox{max width=\hsize}{
    \input{images/mistrain_location.tikz}
 }
   \caption{
        A branch can be mistrained either by the victim process (\emph{same-address-space}) or by an attacker-controlled process (\emph{cross-address-space}).
        Mistraining can be achieved either using the vulnerable branch itself (\emph{in-place}) or a branch at a congruent virtual address (\emph{out-of-place}).}
    \label{fig:mistrain_location}
    \vspace{-0.2cm}
\end{figure}

\paragraph{Systematization of Mistraining Strategies.}
We now propose a second-level classification scheme for Spectre variants that abuse history-based branch prediction (\ie all of the above except \SpectreSTL).
These \Spectre variants first go through a preparatory phase (\cf \cref{fig:transient-flow}) where the microarchitectural branch predictor state is \enquote{poisoned} to cause intentional misspeculation of a particular victim branch.
Since branch prediction buffers in modern CPUs~\cite{Kocher2019spectre,Fog2016} are commonly indexed based on the virtual address of the branch instruction, mistraining can happen either within the same address space or from a different attacker-controlled process.
Furthermore, as illustrated in \cref{fig:mistrain_location}, when only a subset of the virtual address is used in the prediction, mistraining can be achieved using a branch instruction at a congruent virtual address.
We thus enhance the field of \Spectre-type branch poisoning attacks with 4 distinct mistraining strategies:
\begin{compactenum}
    \item Executing the victim branch in the victim process (\emph{same-address-space in-place}).
    \item Executing a congruent branch in the victim process (\emph{same-address-space out-of-place}).
    \item Executing a shadow branch in a different process (\emph{cross-address-space in-place}).
    \item Executing a congruent branch in a different process (\emph{cross-address-space out-of-place}).
\end{compactenum}
In current literature~\cite{Kocher2019spectre,Kiriansky2018speculative,Chen2018SGXpectre,ARMSpecAnalysis_whitepaper}, several of the above branch poisoning strategies have been overlooked for different \Spectre variants.
We summarize the results of an assessment of vulnerabilities under mistraining strategies in \cref{tab:spectre_attack_vulnerability_assessment_pht}.
Our systematization thus reveals clear blind spots that allow an attacker to mistrain branch predictors in previously unknown ways.
As explained further, depending on the adversary's capabilities (\eg in-process, sandboxed, remote, enclave, etc.) these previously unknown mistraining strategies may lead to new attacks and/or bypass existing defenses.

\subsection{\SpectrePHT (Input Validation Bypass)}
\label{sec:spectre:v1}

\paragraph{Microarchitectural Element.}
Kocher~\etal\cite{Kocher2019spectre} first introduced \Spectre\VariantOne, an attack that poisons the Pattern History Table (PHT) to mispredict the direction (taken or not-taken) of conditional branches.
Depending on the underlying microarchitecture, the PHT is accessed based on a combination of virtual address bits of the branch instruction plus a hidden Branch History Buffer (BHB) that accumulates global behavior for the last \emph{N} branches on the same physical core~\cite{Fog2016,Evtyushkin2018BranchScope}

\paragraph{Reading Out-of-Bounds.}
Conditional branches are commonly used by programmers and/or compilers to maintain memory safety invariants at runtime.
For example, consider the following code snippet for bounds checking~\cite{Kocher2019spectre}:

\lstinline[language=C,style=customc]|if (x < len(array1)) { y = array2[array1[x] * 4096]; }|

\noindent
At the architectural level, this program clearly ensures that the index variable \texttt{x} always lies within the bounds of the fixed-length buffer \texttt{array1}.
However, after repeatedly supplying valid values of \texttt{x}, the PHT will reliably predict that this branch evaluates to true.
When the adversary now supplies an invalid index \texttt{x}, the CPU continues along a mispredicted path and transiently performs an out-of-bounds memory access.
The above code snippet features an explicit example of a \enquote{leak gadget} that may act as a microarchitectural covert channel:
depending on the out-of-bounds value being read, the transient instructions load another memory page belonging to \texttt{array2} into the cache.

\paragraph{Writing Out-of-Bounds.}
Kiriansky and Waldspurger~\cite{Kiriansky2018speculative} showed that transient writes are also possible by following the same principle.
Consider the following code line:

\lstinline[language=C,style=customc]|if (x < len(array)) { array[x] = value; }|

\noindent
After mistraining the PHT component, attackers controlling the untrusted index \texttt{x} can transiently write to arbitrary out-of-bounds addresses.
This creates a transient buffer overflow, allowing the attacker to bypass both type and memory safety.
Ultimately, when repurposing traditional techniques from return-oriented programming~\cite{Shacham2007} attacks, adversaries may even gain arbitrary code execution
in the transient domain by overwriting return addresses or code pointers.

\begin{table}[t]
	\setlength{\aboverulesep}{0pt}
	\setlength{\belowrulesep}{0pt}
	\caption{Spectre-type attacks performed in-place, out-of-place, same-address-space (\ie intra-process), or cross-address-space (\ie cross-process).}\label{tab:spectre_attack_vulnerability_assessment_pht}
\begin{center}
\vspace{-0.45cm}
\adjustbox{max width=\columnwidth}{\small{
		\setlength\tabcolsep{1.5pt}
		\begin{tabular}{cc lllll}
		\multicolumn{3}{r}{\diagbox{\textbf{Method}}{\textbf{Attack}}} \quad & \rotatebox{30}{\SpectrePHT} \hspace{-20em} & \rotatebox{30}{\SpectreBTB} \hspace{-20em}  & \rotatebox{30}{\SpectreRSB} \hspace{-20em}  & \rotatebox{30}{\SpectreSTL} \hspace{0em} \\
			\toprule \vspace{-0.35cm}\\
			\multirow{4}{*}{Intel} & \multirow{2}{*}{intra-process} & in-place	& \circletfill~\cite{Kocher2019spectre, Kiriansky2018speculative} 	& \starletfill & \circletfill~\cite{Maisuradze2018spectre5} 	& \circletfill~\cite{Horn2018spectre4} \\
			& & out-of-place	& \starletfill 	& \circletfill~\cite{Chen2018SGXpectre} & \circletfill~\cite{Maisuradze2018spectre5, Koruyeh2018spectre5} 	& \circlet \\
			\cdashline{2-7}
			& \multirow{2}{*}{cross-process} & in-place	& \starletfill 	& \circletfill~\cite{Kocher2019spectre,Chen2018SGXpectre} & \circletfill~\cite{Maisuradze2018spectre5, Koruyeh2018spectre5} & \circlet \\
			& & out-of-place	& \starletfill 	& \circletfill~\cite{Kocher2019spectre} & \circletfill~\cite{Koruyeh2018spectre5}	& \circlet \\
 			\midrule
			\multirow{4}{*}{ARM} & \multirow{2}{*}{intra-process} & in-place	& \circletfill~\cite{Kocher2019spectre, Kiriansky2018speculative} & \starletfill & \circletfill~\cite{ARMSpecAnalysis_whitepaper} 	& \circletfill~\cite{ARMSpecAnalysis_whitepaper} \\
			& & out-of-place	& \starletfill 	& \starlet &  \circletfill~\cite{ARMSpecAnalysis_whitepaper}	& \circlet \\
 			\cdashline{2-7}
			& \multirow{2}{*}{cross-process} & in-place	& \starletfill 	& \circletfill~\cite{ARMSpecAnalysis_whitepaper,Kocher2019spectre} & \starlet & \circlet \\
			& & out-of-place	& \starletfill 	& \starlet & \starlet 	& \circlet \\
 			\midrule
			\multirow{4}{*}{AMD} & \multirow{2}{*}{intra-process} & in-place	& \circletfill~\cite{Kocher2019spectre} 	& \starletfill &  \starletfill	& \circletfill~\cite{Horn2018spectre4} \\
			& & out-of-place	& \starletfill 	& \starlet &  \starletfill	& \circlet \\
 			\cdashline{2-7}
			& \multirow{2}{*}{cross-process}& in-place	& \starletfill 	& \circletfill~\cite{Kocher2019spectre} &  \starletfill	& \circlet \\
			& & out-of-place	& \starletfill 	& \starlet &  \starletfill	& \circlet \\
			\bottomrule
		\end{tabular}

	}}
\end{center}
\vspace{-0.3cm}
    {\footnotesize
    Symbols indicate whether an attack is possible and known (\circletfill), not possible and known (\circlet), possible and previously unknown or not shown (\starletfill), or tested and did not work and previously unknown or not shown (\starlet).
    All tests performed with no defenses enabled.
    }
    \vspace{-0.3cm}
\end{table}

\paragraph{Overlooked Mistraining Strategies.}
\SpectrePHT attacks so far~\cite{Kocher2019spectre,OKeeffe18sgxspectre,Kiriansky2018speculative} rely on a same-address-space in-place branch poisoning strategy.
However, our results (\cf~\cref{tab:spectre_attack_vulnerability_assessment_pht}) reveal that the Intel, ARM, and AMD CPUs we tested are vulnerable to all four PHT mistraining strategies.
In this, we are the first to successfully demonstrate \SpectrePHT-style branch misprediction attacks \emph{without prior execution of the victim branch}.
This is an important contribution as it may open up previously unknown attack avenues for restricted adversaries.

Cross-address-space PHT poisoning may, for instance, enable advanced attacks against a privileged daemon process that does not directly accept user input.
Likewise, for Intel SGX technology, remote attestation schemes have been developed~\cite{Shih2017tsgx} to enforce that a victim enclave can only be run exactly once.
This effectively rules out current state-of-the-art SGXSpectre~\cite{OKeeffe18sgxspectre} attacks that repeatedly execute the victim enclave to mistrain the PHT branch predictor.
Our novel out-of-place PHT poisoning strategy, on the other hand, allows us to perform the training phase entirely \emph{outside} the enclave on the same physical core by repeatedly executing a congruent branch in the untrusted enclave host process (\cf \cref{fig:mistrain_location}).

\subsection{\SpectreBTB (Branch Target Injection)}

\paragraph{Microarchitectural Element.}
In \Spectre\VariantTwo~\cite{Kocher2019spectre}, the attacker poisons the Branch Target Buffer (BTB) to steer the transient execution to a mispredicted branch target.
For direct branches, the CPU indexes the BTB using a subset of the virtual address bits of the branch instruction to yield the predicted jump target.
For indirect branches, CPUs use different mechanisms~\cite{Horn2018spectre}, which may take into account global branching history accumulated in the BHB when indexing the BTB.
We refer to both types as \SpectreBTB.

\paragraph{Hijacking Control Flow.}
Contrary to \SpectrePHT, where transient instructions execute along a restricted mispredicted path, \SpectreBTB allows redirecting transient control flow to an arbitrary destination.
Adopting established techniques from return-oriented programming (ROP) attacks~\cite{Shacham2007}, but abusing BTB poisoning instead of application-level vulnerabilities,
selected code \enquote{gadgets} found in the victim address space may be chained together to construct arbitrary transient instruction sequences.
Hence, where the success of \SpectrePHT critically relies on unintended leakage along the mispredicted code path, ROP-style gadget abuse in \SpectreBTB allows to more directly construct covert channels that expose secrets from the transient domain (\cf \cref{fig:transient-flow}).
We discuss gadget types in more detail in \cref{sec:prevalence}.

\paragraph{Overlooked Mistraining Strategies.}
\SpectreBTB was initially demonstrated on Intel, AMD, and ARM CPUs using a cross-address-space in-place mistraining strategy~\cite{Kocher2019spectre}.
With SGXPectre~\cite{Chen2018SGXpectre}, Chen~\etal extracted secrets from Intel SGX enclaves using either a cross-address-space in-place or same-address-space out-of-place BTB poisoning strategy.
We experimentally reproduced these mistraining strategies  through a systematic evaluation presented in \cref{tab:spectre_attack_vulnerability_assessment_pht}.
On AMD and ARM, we could not demonstrate out-of-place BTB poisoning.
Possibly, these CPUs use an unknown (sub)set of virtual address bits or a function of bits which we were not able to reverse engineer.
We encourage others to investigate whether a different (sub)set of virtual address bits is required to enable the attack.

To the best of our knowledge, we are the first to recognize that \SpectreBTB mistraining can also proceed by \emph{repeatedly executing the vulnerable indirect branch with valid inputs}.
Much like \SpectrePHT, such same-address-space in-place BTB (Spectre-BTB-SA-IP) poisoning abuses the victim's own execution to mistrain the underlying branch target predictor.
Hence, as an important contribution to understanding attack surface and defenses, in-place mistraining \emph{within} the victim domain may allow bypassing widely deployed mitigations~\cite{AMDSpecAnalysis,IntelMitigations} that flush and/or partition the BTB before entering the victim.
Since the branch destination address is now determined by the victim code and not under the direct control of the attacker, however, \SpectreBTB-SA-IP cannot offer the full power of arbitrary transient control flow redirection.
Yet, in higher-level languages like C++ that commonly rely on indirect branches to implement polymorph abstractions, \SpectreBTB-SA-IP may lead to subtle \enquote{speculative type confusion} vulnerabilities.
For example, a victim that repeatedly executes a virtual function call with an object of \texttt{TypeA} may inadvertently mistrain the branch target predictor to cause
misspeculation when finally executing the virtual function call with an object of another \texttt{TypeB}.

\subsection{\SpectreRSB (Return Address Injection)}

\paragraph{Microarchitectural Element.}
Maisuradze and Rossow~\cite{Maisuradze2018spectre5} and Koruyeh~\etal\cite{Koruyeh2018spectre5} introduced a \Spectre variant that exploits the Return Stack Buffer (RSB).
The RSB is a small per-core microarchitectural buffer that stores the virtual addresses following the $N$ most recent \inst{call} instructions.
When encountering a \inst{ret} instruction, the CPU pops the topmost element from the RSB to predict the return flow.

\paragraph{Hijacking Return Flow.}
Misspeculation arises whenever the RSB layout diverges from the actual return addresses on the software stack.
Such disparity for instance naturally occurs when restoring kernel/enclave/user stack pointers upon protection domain switches.
Furthermore, same-address-space adversaries may explicitly overwrite return addresses on the software stack, or transiently execute \inst{call} instructions which update the RSB without committing architectural effects~\cite{Koruyeh2018spectre5}.
This may allow untrusted code executing in a sandbox to transiently divert return control flow to interesting code gadgets outside of the sandboxed environment.

Due to the fixed-size nature of the RSB, a special case of misspeculation occurs for deeply nested function calls~\cite{Koruyeh2018spectre5,Maisuradze2018spectre5}.
Since the RSB can only store return addresses for the $N$ most recent calls, an underfill occurs when the software stack is unrolled.
In this case, the RSB can no longer provide accurate predictions.
Starting from Skylake, Intel CPUs use the BTB as a fallback~\cite{Fog2016,Koruyeh2018spectre5}, thus allowing \SpectreBTB-style attacks triggered by \inst{ret} instructions.

\paragraph{Overlooked Mistraining Strategies.}
\SpectreRSB has been demonstrated with all four mistraining strategies, but only on Intel~\cite{Maisuradze2018spectre5,Koruyeh2018spectre5}.
Our experimental results presented in \cref{tab:spectre_attack_vulnerability_assessment_pht} generalize these strategies to AMD CPUs.
Furthermore, in line with ARM's own analysis~\cite{ARMSpecAnalysis_whitepaper}, we successfully poisoned RSB entries within the same-address-space but did not observe any cross-address-space leakage on ARM CPUs.
We expect this may be a limitation of our current proof-of-concept code and encourage others to investigate this further.

\subsection{\SpectreSTL (Speculative Store Bypass)}
\label{sec:spectre-stl}

\paragraph{Microarchitectural Element.}
Speculation in modern CPUs is not restricted to control flow but also includes predicting dependencies in the data flow.
A common type of Store To Load (STL) dependencies require that a memory load shall not be executed before all preceding stores that write to the same location
have completed.
However, even before the addresses of all prior stores in the pipeline are known, the CPUs' memory disambiguator~\cite{Intel_opt,AMDssbd_whitepaper,Islam2019Spoiler} may predict which loads can already be executed speculatively.

When the disambiguator predicts that a load does not have a dependency on a prior store, the load reads data from the L1 data cache.
When the addresses of all prior stores are known, the prediction is verified.
If any overlap is found, the load and all following instructions are re-executed.

\paragraph{Reading Stale Values.}
Horn~\cite{Horn2018spectre4} showed how mispredictions by the memory disambiguator could be abused to speculatively bypass store instructions.
Like previous attacks, \SpectreSTL adversaries rely on an appropriate transient instruction sequence to leak unsanitized stale values via a microarchitectural covert channel.
Furthermore, operating on stale pointer values may speculatively break type and memory safety guarantees in the transient execution domain~\cite{Horn2018spectre4}.


\begin{table}[t]
	\setlength{\aboverulesep}{0pt}
	\setlength{\belowrulesep}{0pt}
  \caption{Demonstrated Meltdown-type (MD) attacks.}
\begin{center}
\vspace{-0.2cm}
\adjustbox{max width=\columnwidth}{\small{
		\setlength\tabcolsep{1.5pt}
		\begin{tabular}{l llllll:lllllll}
			\textbf{Attack} &\,\, & \rotatebox{45}{\#GP} \hspace{-20em} & \rotatebox{45}{\#NM} \hspace{-20em}  & \rotatebox{45}{\#BR} \hspace{-20em} & \rotatebox{45}{\#PF} \hspace{-20em} & \, \, \hspace{-5em} & \, \, & \rotatebox{45}{U/S} \hspace{-20em} & \rotatebox{45}{P} \hspace{-20em} & \rotatebox{45}{R/W} \hspace{-20em} & \rotatebox{45}{RSVD} \hspace{-20em} & \rotatebox{45}{XD} \hspace{-20em} & \rotatebox{45}{PK} \hspace{-20em} \\
			\toprule\vspace{-10pt} \\
      MD-GP (\MeltdownA)~\cite{ARMSpecAnalysis}           & & \circletfill & \cmarkempty  & \cmarkempty  & \cmarkempty  & \\
      MD-NM (\LazyFP)~\cite{Stecklina2018LazyFP}          & & \cmarkempty  & \circletfill & \cmarkempty  & \cmarkempty  & \\
			MD-BR                                 & & \cmarkempty  & \cmarkempty  & \circletfill & \cmarkempty  & \\
      MD-US (\Meltdown)~\cite{Lipp2018meltdown}           & & \cmarkempty  & \cmarkempty  & \cmarkempty  & \circletfill & & & \cmarkfull  & \cmarkempty & \cmarkempty &  \cmarkempty & \cmarkempty  &  \cmarkempty \\
      MD-P (\Foreshadow)~\cite{Vanbulck2018,Weisse2018foreshadowNG}             & & \cmarkempty  & \cmarkempty  & \cmarkempty  & \circletfill & & & \cmarkempty & \cmarkfull  & \cmarkempty &  \cmarkfull  & \cmarkempty &  \cmarkempty \\
      MD-RW (\VariantOneTwo)~\cite{Kiriansky2018speculative} & & \cmarkempty  & \cmarkempty  & \cmarkempty  & \circletfill & & & \cmarkempty & \cmarkempty & \cmarkfull  &  \cmarkempty & \cmarkempty &  \cmarkempty \\
			MD-PK                                 & & \cmarkempty  & \cmarkempty  & \cmarkempty  & \circletfill & & & \cmarkempty & \cmarkempty & \cmarkempty &  \cmarkempty & \cmarkempty &  \cmarkfull  \\
			\bottomrule
		\end{tabular}
	}}
	\end{center}
  \vspace{-0.3cm}
  {\footnotesize
    Symbols (\circletfill{} or \circlet) indicate whether an exception type (left) or permission bit (right) is exploited.
    Systematic names are derived from what is exploited.
  }
  \label{tab:meltdown-fault-categorization}\label{tab:meltdown-pte-sub-categorization}
    \vspace{-0.5cm}
\end{table}

\section{\Meltdown-type Attacks}
\label{sec:meltdown}
This section overviews \MD-type attacks, and presents a classification scheme that led to the discovery of two previously overlooked \MD variants (\cf \cref{fig:tree}).
Importantly, where \Spectre-type attacks exploit (branch) misprediction events to trigger transient execution, \MD-type attacks rely on transient instructions following a CPU exception.
Essentially, \MD exploits that exceptions are only raised (\ie become architecturally visible) upon the retirement of the faulting instruction.
In some microarchitectures, this property allows transient instructions ahead in the pipeline to compute on unauthorized results of the instruction that is about to suffer a fault.
The CPU's in-order instruction retirement mechanism takes care to discard any architectural effects of such computations, but as with the \Spectre-type attacks above, secrets may leak through microarchitectural covert channels.

\begin{table}[t]
	\setlength{\aboverulesep}{0pt}
	\setlength{\belowrulesep}{0pt}
	\caption{Secrets recoverable via Meltdown-type attacks and whether they cross the current privilege level (CPL).}\label{tab:meltdown-leak-categorization}
\begin{center}
\vspace{-0.6cm}
\adjustbox{max width=\columnwidth}{\small{
		\setlength\tabcolsep{1.5pt}
		\begin{tabular}{l lllllll}
			\diagbox{\textbf{Attack}}{\textbf{Leaks}} &\,\, & & \rotatebox{30}{\hspace{-2em}Memory} \hspace{-20em} & \rotatebox{30}{\hspace{-2em}Cache} \hspace{-20em} & \rotatebox{30}{\hspace{-2em}Register} \hspace{-20em} &\rotatebox{30}{} \hspace{1em} & \rotatebox{30}{\hspace{-2em}Cross-CPL} \hspace{-2em} \\
			\toprule \vspace{-0.3cm}\\
			Meltdown-US (\MD)~\cite{Lipp2018meltdown}           & & \circletfill   & \circletfill & \cmarkempty  & \phantom{--}\qquad  & \cmark \\
			Meltdown-P (\Foreshadow-NG)~\cite{Weisse2018foreshadowNG} & & \circlet & \circletfill & \cmarkempty  &                     & \cmark \\
			Meltdown-P (\Foreshadow-SGX)~\cite{Vanbulck2018} & & \circletfillhl & \circletfill & \circletfillhl &                     & \cmark \\
			Meltdown-GP (\MeltdownA)~\cite{ARMSpecAnalysis}           & & \cmarkempty    & \cmarkempty  & \circletfill &                     & \cmark \\
			Meltdown-NM (\LazyFP)~\cite{Stecklina2018LazyFP}          & & \cmarkempty    & \cmarkempty  & \circletfill &                     & \cmark \\
			Meltdown-RW (\VariantOneTwo)~\cite{Kiriansky2018speculative} & & \circletfill   & \circletfill & \cmarkempty  &                     & \xmark \\
			Meltdown-PK                                 & & \starlet       & \starletfill & \starlet     &                     & \xmark \\
			Meltdown-BR                                 & & \starletfill   & \starletfill & \starlet     &                     & \xmark \\
			\bottomrule
		\end{tabular}
	}}
	\end{center}
\vspace{-0.3cm}
  {\footnotesize
  Symbols indicate whether an attack crosses a processor privilege level (\cmark) or not (\xmark),
  whether it can leak secrets from a buffer (\circletfill),
  only with additional steps (\circletfillhl), or not at all (\circlet).
  Respectively (\starletfill\space vs.\ \starlet) if first shown in this work.
  }
    \vspace{-1.0cm}
\end{table}

\paragraph{Systematization of \MD Variants.}
We introduce a classification for \Meltdown-type attacks in two dimensions.
In the first level, we categorize attacks based on the exception that causes transient execution.
Following Intel's~\cite{Intel_vol3} classification of exceptions as \emph{faults}, \emph{traps}, or \emph{aborts}, we observed that \MD-type attacks so far have exploited faults, but not traps or aborts.
The CPU generates faults if a correctable error has occurred, \ie they allow the program to continue after it has been resolved.
Traps are reported immediately after the execution of the instruction, \ie when the instruction retires and becomes architecturally visible.
Aborts report some unrecoverable error and do not allow a restart of the task that caused the abort.

In the second level, for page faults (\PF), we further categorize based on page-table entry protection bits (\cf \cref{tab:meltdown-fault-categorization}).
We also categorize attacks based on which storage locations can be reached, and whether it crosses a privilege boundary (\cf \cref{tab:meltdown-leak-categorization}).
Through this systematization, we discovered several previously unknown \MD variants that exploit different exception types as well as page-table protection bits, including two exploitable ones.
Our systematic analysis furthermore resulted in the first demonstration of exploitable Meltdown-type delayed exception handling effects on AMD CPUs.

\subsection{\MeltdownUS (Supervisor-only Bypass)}\label{sec:meltdown-meltdown}
Modern CPUs commonly feature a \enquote{user/supervisor} page-table attribute to denote a virtual memory page as belonging to the OS kernel.
The original \Meltdown attack~\cite{Lipp2018meltdown} reads kernel memory from user space on CPUs that do \emph{not} transiently enforce the user/supervisor flag.
In the trigger phase (\cf \cref{fig:transient-flow}) an unauthorized kernel address is dereferenced, which eventually causes a page fault.
Before the fault becomes architecturally visible, however, the attacker executes a transient instruction sequence that for instance accesses a cache line based on the privileged data read by the trigger instruction.
In the final phase, after the exception has been raised, the privileged data is reconstructed at the receiving end of the covert channel (\eg \FlushReload).

The attacks bandwidth can be improved by suppressing exceptions through transaction memory CPU features such as Intel TSX~\cite{Intel_vol3}, exception handling~\cite{Lipp2018meltdown}, or hiding it in another transient execution~\cite{Horn2018spectre, Lipp2018meltdown}.
By iterating byte-by-byte over the kernel space and suppressing or handling exceptions, an attacker can dump the entire kernel.
This includes the entire physical memory if the operating system has a direct physical map in the kernel.
While extraction rates are significantly higher when the kernel data resides in the CPU cache, \MD has even been shown to successfully extract uncached data from memory~\cite{Lipp2018meltdown}.

\subsection{\MeltdownP (Virtual Translation Bypass)}\label{sec:meltdown-foreshadow}
\paragraph{Foreshadow.}
Van~Bulck~\etal\cite{Vanbulck2018} presented \Foreshadow, a \Meltdown-type attack targeting Intel SGX technology~\cite{Intel_SGX}.
Unauthorized accesses to enclave memory usually do not raise a \PF exception but are instead silently replaced with abort page dummy values (\cf \cref{sec:meltdown-defense-d2}).
In the absence of a fault, plain \MD cannot be mounted against SGX enclaves.
To overcome this limitation, a \FS attacker clears the \enquote{present} bit in the page-table entry mapping the enclave secret, ensuring that a \PF will be raised for subsequent accesses.
Analogous to \MeltdownUS, the adversary now proceeds with a transient instruction sequence to leak the secret (\eg through a \FlushReload covert channel).

Intel~\cite{Intel2018l1tf} named \textit{L1 Terminal Fault} (L1TF) as the root cause behind \FS.
A terminal fault occurs when accessing a page-table entry with either the present bit cleared or a ``reserved'' bit set.
In such cases, the CPU immediately aborts address translation.
However, since the L1 data cache is indexed in parallel to address translation, the page table entry's physical address field (\ie frame number) may still be passed to the L1 cache.
Any data present in L1 and tagged with that physical address will now be forwarded to the transient execution, regardless of access permissions.

Although \MeltdownP-type leakage is restricted to the L1 data cache, the original \FS~\cite{Vanbulck2018} attack showed how SGX's secure page swapping mechanism might first be abused to prefetch arbitrary enclave pages into the L1 cache, including even CPU registers stored on interrupt.
This highlights that SGX's privileged adversary model considerably amplifies the transient execution attack surface.

\paragraph{\ForeshadowNG.}\label{sec:meltdown-foreshadow-ng}
\ForeshadowNG~\cite{Weisse2018foreshadowNG} generalizes \Foreshadow from the attack on SGX enclaves to bypass operating system or hypervisor isolation.
The generalization builds on the observation that the physical frame number in a page-table entry is sometimes under direct or indirect control of an adversary.
For instance, when swapping pages to disk, the kernel is free to use all but the present bit to store metadata (\eg the offset on the swap partition).
However, if this offset is a valid physical address, any cached memory at that location leaks to an unprivileged \FS-OS attacker.

Even worse is the Foreshadow-VMM variant, which allows an untrusted virtual machine, controlling guest-physical addresses, to extract the host machine's entire L1 data cache (including data belonging to the hypervisor or other virtual machines).
The underlying problem is that a terminal fault in the guest page-tables early-outs the address translation process, such that guest-physical addresses are erroneously passed to the L1 data cache, without first being translated into a proper host physical address~\cite{Intel2018l1tf}.

\subsection{\MeltdownGP (System Register Bypass)}\label{sec:meltdown-meltdown3a}
\MeltdownGP (named initially \MeltdownA)~\cite{Intel2018m3a} allows an attacker to read privileged system registers.
It was first discovered and published by ARM~\cite{ARMSpecAnalysis} and subsequently Intel~\cite{Intel2018white_paper} determined that their CPUs are also susceptible to the attack.
Unauthorized access to privileged system registers (\eg via \inst{rdmsr}) raises a \textit{general protection} fault (\GP).
Similar to previous \MD-type attacks, however, the attack exploits that the transient execution following the faulting instruction can still compute on the unauthorized data, and leak the system register contents through a microarchitectural covert channel (\eg \FlushReload).

\subsection{\MeltdownNM (FPU Register Bypass)}\label{sec:meltdown-lazyfp}
During a context switch, the OS has to save all the registers, including the floating point unit (FPU) and SIMD registers.
These latter registers are large and saving them would slow down context switches.
Therefore, CPUs allow for a lazy state switch, meaning that instead of saving the registers, the FPU is simply marked as ``not available''.
The first FPU instruction issued after the FPU was marked as ``not available'' causes a \textit{device-not-available} (\NM) exception, allowing the OS to save the FPU state of previous execution context before marking the FPU as available again.

Stecklina and Prescher~\cite{Stecklina2018LazyFP} propose an attack on the above lazy state switch mechanism.
The attack consists of three steps.
In the first step, a victim performs operations loading data into the FPU registers.
Then, in the second step, the CPU switches to the attacker and marks the FPU as ``not available''.
The attacker now issues an instruction that uses the FPU, which generates an \NM fault.
Before the faulting instruction retires, however, the CPU has already transiently executed the following instructions using data from the previous context.
As such, analogous to previous \MD-type attacks, a malicious transient instruction sequence following the faulting instruction can encode the unauthorized FPU register contents through a microarchitectural covert channel (\eg \FlushReload).

\subsection{\MeltdownRW (Read-only Bypass)}\label{sec:meltdown-rw}
Where the above attacks~\cite{Lipp2018meltdown,Vanbulck2018,ARMSpecAnalysis,Stecklina2018LazyFP}
focussed on stealing information across privilege levels, Kiriansky and Waldspurger~\cite{Kiriansky2018speculative} presented the first \MD-type attack that bypasses page-table based access rights \emph{within} the current privilege level.
Specifically, they showed that transient execution does not respect the \enquote{read/write} page-table attribute.
The ability to transiently overwrite read-only data within the current privilege level can bypass software-based sandboxes which rely on hardware enforcement of read-only memory.

Confusingly, the above \MeltdownRW attack was originally named ``Spectre Variant 1.2''~\cite{Kiriansky2018speculative} as the authors followed a Spectre-centric naming scheme.
Our systematization revealed, however, that the transient cause exploited above is a \PF exception.
Hence, this attack is of \MD-type, but \emph{not} a variant of \Spectre.

\subsection{\MeltdownPK (Protection Key Bypass)}\label{sec:meltdown-pk}
Intel Skylake-SP server CPUs support memory-protection keys for user space (PKU)~\cite{Intel2018pku}.
This feature allows processes to change the access permissions of a page directly from user space, \ie without requiring a syscall/hypercall.
Thus, with PKU, user-space applications can implement efficient hardware-enforced isolation of trusted parts~\cite{VahldiekOberwagner2018,Hedayati2018}.

We present a novel \MeltdownPK attack to bypass both read and write isolation provided by PKU.
\MeltdownPK works if an attacker has code execution in the containing process, even if the attacker cannot execute the \inst{wrpkru} instruction (\eg blacklisting).
Moreover, in contrast to cross-privilege level \MD attack variants, there is no software workaround.
According to Intel~\cite{Intel2018sok_mitigation}, \MeltdownPK can be mitigated using address space isolation.
Recent Meltdown-resistant Intel processors enumerating \rdclno plus PKU support furthermore mitigate \MeltdownPK in silicon.
With those mitigations, the memory addresses that might be revealed by transient execution attacks can be limited.

\paragraph{Experimental Results.}
We tested \MeltdownPK on an Amazon EC2 C5 instance running Ubuntu 18.04 with PKU support.
We created a memory mapping and used PKU to remove both read and write access.
As expected, protected memory accesses produce a \PF.
However, our proof-of-concept manages to leak the data via an adversarial transient instruction sequence with a \FlushReload covert channel.

\subsection{\MeltdownBR (Bounds Check Bypass)}
\label{sec:meltdown-br}

To facilitate efficient software instrumentation, x86 CPUs come with dedicated hardware instructions that raise a \textit{bound-range-exceeded} exception (\BR) when encountering out-of-bound array indices.
The IA-32 ISA, for instance, defines a \inst{bound} opcode for this purpose.
While the \inst{bound} instruction was omitted in the subsequent x86-64 ISA, modern Intel CPUs ship with Memory Protection eXtensions (MPX) for efficient array bounds checking.

Our systematic evaluation revealed that Meltdown-type effects of the \BR exception had not been thoroughly investigated yet.
Specifically, Intel's analysis~\cite{IntelMitigations} only briefly mentions MPX-based bounds check bypass as a possibility, and recent defensive work by Dong~\etal\cite{Dong2018Spectres} highlights the need to introduce a memory \inst{lfence} after MPX bounds check instructions.
They classify this as a Spectre-type attack, implying  that the \inst{lfence} is needed to prevent the branch predictor from speculating on the outcome of the bounds check.
According to Oleksenko~\etal\cite{Oleksenko2017mpx}, neither \inst{bndcl} nor \inst{bndcu} exert pressure on the branch predictor, indicating that there is no prediction happening.
Based on that, we argue that the classification as a \Spectre-type attack is misleading as no prediction is involved.
The observation by Dong~\etal\cite{Dong2018Spectres} indeed does not shed light on the \BR exception as the root cause for the MPX bounds check bypass, and they do not consider IA32 \inst{bound} protection at all.
Similar to \SpectrePHT, \MeltdownBR is a bounds check bypass, but instead of mistraining a predictor it exploits the lazy handling of the raised \BR exception.

\paragraph{Experimental Results.}
We introduce the \MeltdownBR attack which exploits transient execution following a \BR exception to encode out-of-bounds secrets that are never architecturally visible.
As such, \MeltdownBR is an exception-driven alternative for \SpectrePHT.
Our proofs-of-concept demonstrate out-of-bounds leakage through a \FlushReload covert channel for an array index safeguarded by either IA32 \inst{bound} (Intel, AMD), or state-of-the-art MPX protection (Intel-only).
For Intel, we ran the attacks on a Skylake i5-6200U CPU with MPX support, and for AMD we evaluated both an E2-2000 and a Ryzen Threadripper 1920X.
This is the first experiment demonstrating a Meltdown-type transient execution attack exploiting delayed exception handling on AMD CPUs~\cite{AMDSpecAnalysis,Lipp2018meltdown}.

\begin{table}[t]
	\setlength{\aboverulesep}{0pt}
	\setlength{\belowrulesep}{0pt}
    \newcommand{\na}{\lineh}
  \caption{CPU vendors vulnerable to \Meltdown (MD).}\label{tab:meltdown-vulnerable-vendor}
\begin{center}
\vspace{-0.5cm}
\adjustbox{max width=\columnwidth}{\small{
		\setlength\tabcolsep{1.5pt}
		\begin{tabular}{r lllllllllllllll}
			\diagbox{\textbf{Vendor}}{\textbf{Attack}} &\,\, &
                    \rotatebox{40}{MD-US~\cite{Lipp2018meltdown}} \hspace{-20em} &
                    \rotatebox{40}{MD-P~\cite{Vanbulck2018,Weisse2018foreshadowNG}} \hspace{-20em} &
                    \rotatebox{40}{MD-GP~\cite{ARMSpecAnalysis,Intel2018white_paper}} \hspace{-20em} &
                    \rotatebox{40}{MD-NM~\cite{Stecklina2018LazyFP}} \hspace{-20em} &
                    \rotatebox{40}{MD-RW~\cite{Kiriansky2018speculative}} \hspace{-20em} &
                    \rotatebox{40}{MD-PK} \hspace{-20em} &
                    \rotatebox{40}{MD-BR} \hspace{-20em} &
                    \rotatebox{40}{MD-DE} \hspace{-20em} &
                    \rotatebox{40}{MD-AC} \hspace{-20em} &
                    \rotatebox{40}{MD-UD} \hspace{-20em} &
                    \rotatebox{40}{MD-SS} \hspace{-20em} &
                    \rotatebox{40}{MD-XD} \hspace{-20em} &
                    \rotatebox{40}{MD-SM} \hspace{-20em} & \hspace{2.75em} \\
			\toprule \vspace{-0.3cm}\\
			Intel && \circletfill 	& \circletfill 	& \circletfill	 & \circletfill & \circletfill 	& \cstarletfill  & \cstarletfill & \cstarlet  & \cstarlet & \cstarlet  & \cstarlet  & \cstarlet & \cstarlet\\
			ARM   && \circletfill 	& \cmarkempty   & \circletfill 	 & \na	        & \circletfill  & \na            & \na 	         & \cstarlet  & \cstarlet & \cstarlet  & \na        & \cstarlet & \cstarlet\\
			AMD   && \cmarkempty 	& \cmarkempty	& \cmarkempty 	 & \cmarkempty  & \cmarkempty	& \na            & \cstarletfill & \cstarlet  & \cstarlet & \cstarlet  & \cstarlet  & \cstarlet & \cstarlet\\
			\bottomrule
		\end{tabular}
	}}
\end{center}
\vspace{-0.3cm}
\footnotesize{
    Symbols indicate whether at least one CPU model is vulnerable (filled) vs.\ no CPU is known to be vulnerable (empty).
    Glossary: reproduced (\circletfill{} vs.\ \circlet),
              first shown in this paper (\starletfill{} vs.\ \starlet{}),
              not applicable (\na).
    All tests performed without defenses enabled.
}
    \vspace{-0.5cm}
\end{table}

\begin{table*}[t]
	\setlength{\aboverulesep}{0pt}
	\setlength{\belowrulesep}{0pt}
	\caption{Gadget classification according to the attack flow and whether executed by the attacker (\circletfill{}), victim (\circlet{}), or either (\circletfillhl{}).}
    \label{tab:gadget-classification}
\begin{center}
\vspace{-0.4cm}
\adjustbox{max width=\textwidth}{\small
		\setlength\tabcolsep{5pt}
        \newcommand{\dito}{\multicolumn{1}{c}{--- \raisebox{-0.5ex}{''} ---}}
		\begin{tabular}{lllll}
			{\bf Attack} & {\bf 1. Preface} & {\bf 2. Trigger example} & {\bf 3. Transient} & {\bf 5. Reconstruction}  \\
			\toprule \vspace{-0.3cm}\\
			Covert channel~\cite{Yarom2014,Aldaya2018,Schwarz2018netspectre} &
                \circletfillhl{} Flush/Prime/Evict & - & \circletfillhl{} Load/AVX/Port/... & \circletfillhl{} Reload/Probe/Time \\
            \midrule
			Meltdown-US/RW/GP/NM/PK~\cite{Lipp2018meltdown,Kiriansky2018speculative,ARMSpecAnalysis,Stecklina2018LazyFP} &
                \circletfill{} (Exception suppression) & \circletfill{} \inst{mov/rdmsr}/FPU & \circletfill{} Controlled encode & \circletfill{} Exception handling \\
			Meltdown-P~\cite{Vanbulck2018,Weisse2018foreshadowNG} &
                \circlet{} (L1 prefetch) & \circletfill{} \inst{mov} & \circletfill{} Controlled encode & \ \ \& controlled decode \\
      Meltdown-BR & - & \circlet{} \inst{bound/bndclu} & \circlet{} Inadvertent leak &  \it same as above \\
            \midrule
			\SpectrePHT~\cite{Kocher2019spectre} &
                \circletfillhl{} PHT poisoning      & \circlet{} \inst{jz}       & \circlet{} Inadvertent leak & \circletfill{} Controlled decode \\
			\SpectreBTB/RSB~\cite{Kocher2019spectre,Chen2018SGXpectre,Maisuradze2018spectre5,Koruyeh2018spectre5} &
                \circletfillhl{} BTB/RSB poisoning  & \circlet{} \inst{call/jmp/ret} & \circlet{} ROP-style encode & \circletfill{} Controlled decode \\
			\SpectreSTL~\cite{Horn2018spectre4} &
                - & \circlet{} \inst{mov}       & \circlet{} Inadvertent leak & \circletfill{} Controlled decode \\
			NetSpectre~\cite{Schwarz2018netspectre} &
                \circlet{} Thrash/reset             & \circlet{} \inst{jz}       & \circlet{} Inadvertent leak & \circlet{} Inadvertent transmit \\
			\bottomrule
		\end{tabular}
	}
	\end{center}
	\vspace{-0.9cm}
\end{table*}

\subsection{Residual Meltdown (Negative Results)}
We systematically studied transient execution leakage for other, not yet tested exceptions.
In our experiments, we consistently found no traces of transient execution beyond traps or aborts, which leads us to the hypothesis that \Meltdown is only possible with faults (as they can occur at any moment during instruction execution).
Still, the possibility remains that our experiments failed and that they are possible.
\Cref{tab:meltdown-vulnerable-vendor} and \cref{fig:tree} summarize experimental results for fault types tested on Intel, ARM, and AMD.

\paragraph{Division Errors.}
For the divide-by-zero experiment, we leveraged the signed division instruction (\inst{idiv} on x86 and \inst{sdiv} on ARM).
On the ARMs we tested, there is no exception, but the division yields merely zero.
On x86, the division raises a \textit{divide-by-zero} exception (\DE).
Both on the AMD and Intel we tested, the CPU continues with the transient execution after the exception.
In both cases, the result register is set to `0', which is the same result as on the tested ARM.
Thus, according to our experiments \MeltdownDE is not possible, as no real values are leaked.

\paragraph{Supervisor Access.}
Although supervisor mode access prevention (SMAP) raises a page fault (\PF) when accessing user-space memory from the kernel, it seems to be free of any \Meltdown effect in our experiments.
Thus, we were not able to leak any data using \MeltdownSM in our experiments.

\paragraph{Alignment Faults.}
Upon detecting an unaligned memory operand, the CPU may generate an \textit{alignment check} exception (\AC).
In our tests, the results of unaligned memory accesses never reach the transient execution.
We suspect that this is because \AC is generated early-on, even before the operand's virtual address is translated to a physical one.
Hence, our experiments with \MeltdownAC were unsuccessful in showing any leakage.

\paragraph{Segmentation Faults.}
We consistently found that out-of-limit segment accesses never reach transient execution in our experiments.
We suspect that, due to the simplistic IA32 segmentation design, segment limits are validated early-on, and immediately raise a \GP or \SF (\textit{stack-segment fault}) exception, without sending the offending instruction to the ROB.
Therefore, we observed no leakage in our experiments with \MeltdownSS.

\paragraph{Instruction Fetch.}
To yield a complete picture, we investigated \MD-type effects during the instruction fetch and decode phases.
On our test systems, we did not succeed in transiently executing instructions residing in non-executable memory (\ie \MeltdownXD), or following an \textit{invalid opcode} (\UD) exception (\ie \MeltdownUD).
We suspect that exceptions during instruction fetch or decode are immediately handled by the CPU, without first buffering the offending instruction in the ROB.
Moreover, as invalid opcodes have an undefined length, the CPU does not even know where the next instruction starts.
Hence, we suspect that invalid opcodes only leak if the microarchitectural effect is already an effect caused by the invalid opcode itself, not by subsequent transient instructions.


\section{Gadget Analysis and Classification}
\label{sec:prevalence}
We deliberately oriented our attack tree (\cf \cref{fig:tree}) on the microarchitectural root causes of the transient computation, abstracting away from the underlying covert channel and/or code \emph{gadgets} required to carry out the attack successfully.
In this section, we further dissect transient execution attacks by categorizing gadget types in two tiers and overviewing current results on their exploitability in real-world software.

\subsection{Gadget Classification}
\label{sec:gadget-classification}
\paragraph{First-Tier: Execution Phase.}
We define a \enquote{gadget} as a series of instructions executed by either the attacker or the victim.
\cref{tab:gadget-classification} shows how gadget types discussed in literature can be unambiguously assigned to one of the abstract attack phases from \cref{fig:transient-flow}.
New gadgets can be added straightforwardly after determining their execution phase and objective.

Importantly, our classification table highlights that gadget choice largely depends on the attacker's capabilities.
By plugging in different gadget types to compose the required attack phases, an almost boundless spectrum of adversary models can be covered that is only limited by the attacker's capabilities.
For local adversaries with arbitrary code execution (\eg \MeltdownUS~\cite{Lipp2018meltdown}), the gadget functionality can be explicitly implemented by the attacker.
For sandboxed adversaries (\eg \SpectrePHT~\cite{Kocher2019spectre}), on the other hand, much of the gadget functionality has to be provided by \enquote{confused deputy} code executing in the victim domain.
Ultimately, as claimed by Schwarz~\etal\cite{Schwarz2018netspectre},
even fully remote attackers may be able to launch \Spectre attacks
given that sufficient gadgets would be available inside the victim code.

\begin{table}[t]
  \setlength{\aboverulesep}{0pt}
  \setlength{\belowrulesep}{0pt}
  \caption{\SpectrePHT gadget classification and the number of occurrences per gadget type in Linux kernel v5.0.}
  \label{tab:gadget-upper-bound}
\begin{center}
\vspace{-0.4cm}
\adjustbox{max width=\columnwidth}{\small{
    \begin{tabular}{l ll}
      \textbf{Gadget}& \textbf{Example (\SpectrePHT)} & \textbf{\#Occurrences} \\
      \toprule \vspace{-0.3cm}\\
      Prefetch      & \texttt{\footnotesize if(i<LEN\_A)\{a[i];\}} & 172 \\
      Compare       & \texttt{\footnotesize if(i<LEN\_A)\{if(a[i]==k)\{\};\}} & 127 \\
      Index         & \texttt{\footnotesize if(i<LEN\_A)\{y = b[a[i]*x];\}} & 0 \\
      Execute       & \texttt{\footnotesize if(i<LEN\_A)\{a[i](void);\}} & 16 \\
      \bottomrule
    \end{tabular}
  }}
  \end{center}
    \vspace{-0.3cm}
\end{table}

\paragraph{Second-Tier: Transient Leakage.}
During our analysis of the Linux kernel (see ~\cref{sec:prevalence-linux}), we discovered that gadgets required for \SpectrePHT can be further classified in a second tier.
A second tier is required in this case as those gadgets enable different types of attacks.
The first type of gadget we found is called \textit{Prefetch}.
A Prefetch gadget consists of a single array access.
As such it is not able to leak data, but can be used to load data that can then be leaked by another gadget as was demonstrated by \MeltdownP~\cite{Vanbulck2018}.
The second type of gadget, called \textit{Compare}, loads a value like in the Prefetch gadget and then branches on it.
Using a contention channel like execution unit contention~\cite{Aldaya2018PortSmash,Bhattacharyya2019smother} or an AVX channel as claimed by Schwarz~\etal\cite{Schwarz2018netspectre}, an attacker might be able to leak data.
We refer to the third gadget as \textit{Index} gadget and it is the double array access shown by Kocher~\etal\cite{Kocher2019spectre}.
The final gadget type, called \textit{Execute}, allows arbitrary code execution, similar to \SpectreBTB.
In such a gadget, an array is indexed based on an attacker-controlled input and the resulting value is used as a function pointer, allowing an attacker to transiently execute code by accessing the array out-of-bounds.
\cref{tab:gadget-upper-bound} gives examples for all four types.

\subsection{Real-World Software Gadget Prevalence}
\label{sec:prevalence-linux}
While for \MD-type attacks, convincing real-world exploits have been developed to dump arbitrary process~\cite{Lipp2018meltdown} and enclave~\cite{Vanbulck2018} memory, most Spectre-type attacks have so far only been demonstrated in controlled environments.
The most significant barrier to mounting a successful Spectre attack is to find exploitable gadgets in real-world software, which at present remains an important open research question in itself~\cite{Maisuradze2018spectre5,Schwarz2018netspectre}.

\paragraph{Automated Gadget Analysis.}
Since the discovery of transient execution attacks, researchers have tried to develop methods for the automatic analysis of gadgets.
One proposed method is called \text{oo7}~\cite{Wang2017oo7} and uses taint tracking to detect \SpectrePHT Prefetch and Index gadgets.
oo7 first marks all variables that come from an untrusted source as tainted.
If a tainted variable is later on used in a branch, the branch is also tainted.
The tool then reports a possible gadget if a tainted branch is followed by a memory access depending on the tainted variable.
Guarnieri~\etal\cite{Guarnieri2018spectector} mention that oo7 would still flag code locations that were patched with Speculative Load Hardening~\cite{Carruth2018Hardening} as it would still match the vulnerable pattern.

Another approach, called Spectector~\cite{Guarnieri2018spectector}, uses symbolic execution to detect \SpectrePHT gadgets.
It tries to formally prove that a program does not contain any gadgets by tracking all memory accesses and jump targets during execution along all different program paths.
Additionally, it simulates the path of mispredicted branches for a number of steps.
The program is run twice to determine whether it is free of gadgets or not
First, it records a trace of memory accesses when no misspeculation occurs (\ie runs the program in its intended way).
Second, it records a trace of memory accesses with misspeculation of a certain number of instructions.
Spectector then reports a gadget if it detects a mismatch between the two traces.
One problem with the Spectector approach is scalability as it is currently not feasible to symbolically execute large programs.

\begin{figure}
    \includegraphics[width=\hsize]{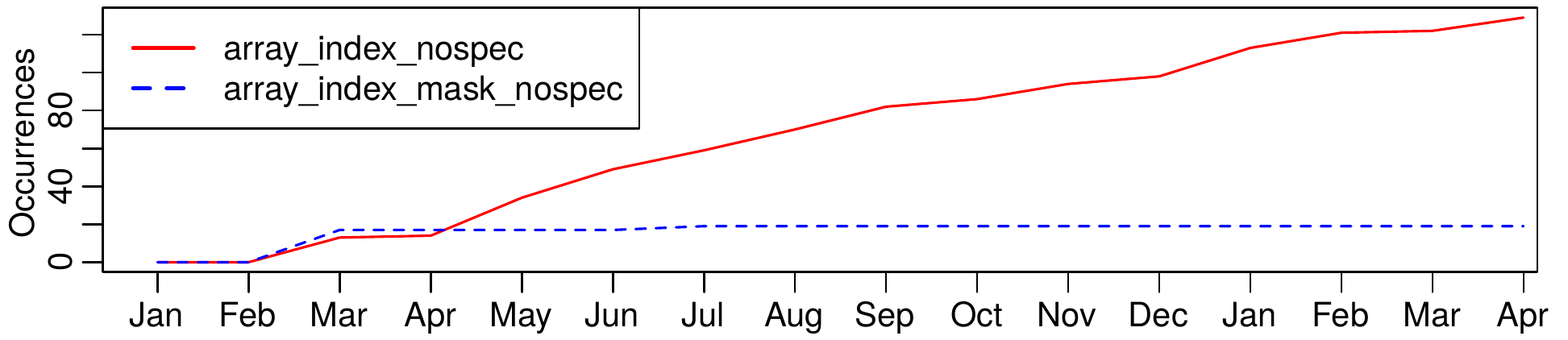}
    \caption{Evolution of \SpectrePHT patches in the Linux kernel over time (2018-2019).}
    \label{fig:linux-graph}
    \vspace{-0.1cm}
\end{figure}

The Linux kernel developers use a different approach.
They extended the Smatch static analysis tool to automatically discover potential Spectre-PHT out-of-bounds access gadgets~\cite{Carpenter2018smatch}.
Specifically, Smatch finds all instances of user-supplied array indices that have not been explicitly hardened.
Unfortunately, Smatch's false positive rate is quite high.
According to Carpenter~\cite{Carpenter2018smatch}, the tool reported 736 gadget candidates in April 2018, whereas the kernel only featured about 15 Spectre-PHT-resistant array indices at that time.
We further investigated this by analyzing the number of occurrences of the newly introduced \inst{array\_index\_nospec} and \inst{array\_index\_mask\_nospec} macros in the Linux kernel per month.
\Cref{fig:linux-graph} shows that the number of \SpectrePHT patches has been continuously increasing over the past year.
This provides further evidence that patching \SpectrePHT gadgets in real-world software is an ongoing effort and that automated detection methods and gadget classification pose an important research challenge.

\paragraph{Academic Review.}
To date, only 5 academic papers have demonstrated Spectre-type gadget exploitation in real-world software~\cite{Kocher2019spectre,Chen2018SGXpectre,Maisuradze2018spectre5,Horn2018spectre4,Bhattacharyya2019smother}.
\Cref{tab:prevalence-gadgets} reveals that they either abuse ROP-style gadgets in larger code bases or more commonly rely on Just-In-Time (JIT) compilation to indirectly provide the vulnerable gadget code.
JIT compilers as commonly used in \eg JavaScript, WebAssembly, or the eBPF Linux kernel interface, create a software-defined sandbox by extending the untrusted attacker-provided code with runtime checks.
However, the attacks in \cref{tab:prevalence-gadgets} demonstrate that such JIT checks can be transiently circumvented to leak memory contents outside of the sandbox.
Furthermore, in the case of \SpectreBTB/RSB, even non-JIT compiled real-world code has been shown to be exploitable when the attacker controls sufficient inputs to the victim application.
Kocher~\etal\cite{Kocher2019spectre} constructed a minimalist proof-of-concept that reads attacker-controlled inputs into registers before calling a function.
Next, they rely on BTB poisoning to redirect transient control flow to a gadget they identified in the Windows \texttt{ntdll} library that allows leaking arbitrary memory from the victim process.
Likewise, Chen~\etal\cite{Chen2018SGXpectre} analyzed various trusted enclave runtimes for Intel SGX and found several instances of vulnerable branches with attacker-controlled input registers, plus numerous exploitable gadgets to which transient control flow may be directed to leak unauthorized enclave memory.
Bhattacharyya~\etal\cite{Bhattacharyya2019smother} analyzed common software libraries that are likely to be linked against a victim program for gadgets.
They were able to find numerous gadgets and were able to exploit one in OpenSSL to leak information.

\begin{table}[t]
	\setlength{\aboverulesep}{0pt}
	\setlength{\belowrulesep}{0pt}
	\caption{Spectre-type attacks on real-world software.}\label{tab:prevalence-gadgets}
\begin{center}
\vspace{-0.4cm}
\adjustbox{max width=\columnwidth}{\small{
		\begin{tabular}{llll}
            {\bf Attack} & {\bf Gadgets} & {\bf JIT} & {\bf Description} \\
			\toprule \vspace{-0.3cm}\\
			\SpectrePHT~\cite{Kocher2019spectre}      & 2   & \cmark        & Chrome Javascript, Linux eBPF \\
			\SpectreBTB~\cite{Kocher2019spectre}      & 2   & \cmark/\xmark & Linux eBPF, Windows \texttt{ntdll} \\
			\SpectreBTB~\cite{Chen2018SGXpectre}      & 336 & \xmark        & SGX SDK Intel/Graphene/Rust \\
      \SpectreBTB~\cite{Bhattacharyya2019smother} & 690 & \xmark      & OpenSSL, glibc, pthread, ... \\
			\SpectreRSB~\cite{Maisuradze2018spectre5} & 1   & \cmark        & Firefox WebAssembly \\
			\SpectreSTL~\cite{Horn2018spectre4}       & 1   & \cmark        & Partial PoC on Linux eBPF \\
			\bottomrule
		\end{tabular}
	}}
	\end{center}
	\vspace{-0.6cm}
\end{table}

\paragraph{Case Study: Linux Kernel.}
To further assess the prevalence of Spectre gadgets in real-world software, we selected the Linux kernel (Version 5.0) as a relevant case study of a major open-source project that underwent numerous Spectre-related security patches over the last year.
We opted for an in-depth analysis of one specific piece of software instead of a breadth-first approach where we do a shallow analysis of multiple pieces of software.
This allowed us to analyse historical data (\ie code locations the kernel developers deemed necessary to protect) that led to the second tier classification discussed in \cref{sec:gadget-classification}.

There are a couple of reasons that make analysis difficult.
The first is that Linux supports many different platforms.
Therefore, particular gadgets are only available in a specific configuration.
The second point is that the number of instructions that can be transiently executed depends on the size of the ROB~\cite{Wang2017oo7}.
As we analyze high-level code, we can only estimate how far ahead the processor can transiently execute.

\cref{tab:gadget-upper-bound} shows the number of occurrences of each gadget type from our second tier classification.
While \cref{fig:linux-graph} shows around 120 occurrences of \inst{array\_index\_nospec}, the number of gadgets in our analysis is higher.
The reason behind that is that multiple arrays are indexed with the same masked index as well as multiple branches that contain a value that was loaded with a potential malicious index.
Our analysis also shows that more dangerous gadgets that either allow more than 1-bit leakage or even arbitrary code execution are not frequently occurring.
Even if one is found, it might still be hard to exploit.
During our analysis, we also discovered that in 13 locations the patch had been reverted, indicating that there is also some confusion with the kernel developers what needs to be fixed.


\section{Defenses}
\label{sec:defenses}
In this section, we discuss proposed defenses in software and hardware for \Spectre and \Meltdown variants.
We propose a classification scheme for defenses based on their attempt to stop leakage, similar to Miller~\cite{Miller2018taxonomy}.
Our work differs from Miller in three points.
First, ours extends to newer transient execution attacks.
Second, we consider Meltdown and Spectre as two problems with different root causes, leading to a different classification.
Third, it helped uncover problems that were not clear with the previous classification.

\noindent
We categorize \Spectre-type defenses into three categories:
\begin{compactenum}
  \item[\textbf{C1}:] Mitigating or reducing the accuracy of covert channels used to extract the secret data.
	\item[\textbf{C2}:] Mitigating or aborting speculation if data is potentially accessible during transient execution.
	\item[\textbf{C3}:] Ensuring that secret data cannot be reached.
\end{compactenum}

\cref{tab:spectre_defenses_categorization} lists proposed defenses against \Spectre-type attacks and assigns them to the category they belong.

\begin{table}[t]
	\setlength{\aboverulesep}{0pt}
	\setlength{\belowrulesep}{0pt}
	\caption{Categorization of Spectre defenses and systematic overview of their microarchitectural target.}\label{tab:spectre_defenses_target_me_element}\label{tab:spectre_defenses_categorization}
\begin{center}
\vspace{-0.5cm}
\adjustbox{max width=\columnwidth}{\small{
		\setlength\tabcolsep{1.5pt}
		\begin{tabular}{r r lllllllllllllllllllllll}
      & \textbf{Defense} &\,\, & \rotatebox{60}{InvisiSpec} \hspace{-20em} & \rotatebox{60}{SafeSpec} \hspace{-20em} & \rotatebox{60}{DAWG} \hspace{-20em} & \rotatebox{60}{Taint Tracking} \hspace{-20em} & \rotatebox{60}{Timer Reduction} \hspace{-20em} & \, & \rotatebox{60}{RSB Stuffing} \hspace{-20em} & \rotatebox{60}{Retpoline} \hspace{-20em} &  \rotatebox{60}{SLH} \hspace{-20em} & \rotatebox{60}{YSNB} \hspace{-20em} & \rotatebox{60}{IBRS} \hspace{-20em} & \rotatebox{60}{STIBP} \hspace{-20em} & \rotatebox{60}{IBPB} \hspace{-20em} & \rotatebox{60}{Serialization} \hspace{-20em} & \rotatebox{60}{Sloth} \hspace{-20em} & \rotatebox{60}{SSBD/SSBB} \hspace{-20em} & \, & \rotatebox{60}{Poison Value} \hspace{-20em} & \rotatebox{60}{Index Masking} \hspace{-20em} & \rotatebox{60}{Site Isolation} \hspace{-20em} & \hspace{2em}  \\
			\toprule \vspace{-0.3cm}\\
\multirow{7}{*}{\rotatebox{90}{\resizebox{3.5cm}{!}{\textbf{Microarchitectural Element}}}} &			Cache &		 		& \circletfill 	& \circletfill	& \circletfill 	& \cmarkempty & \multicolumn{1}{c}{\circlet \phantom{*}} & & \cmarkempty 	& \cmarkempty 	& \cmarkempty 	& \cmarkempty 	&  \cmarkempty 	& \cmarkempty 	&  \cmarkempty 	& \cmarkfull 	& \cmarkempty 	& \multicolumn{1}{c}{\circlet \phantom{*}} &  	&  \cmarkempty & \cmarkempty & \cmarkempty \\
&			TLB & 			& \circletfillhl 	& \circletfill 	& \circletfillhl & \cmarkempty & \multicolumn{1}{c:}{\circlet \phantom{*}} & 	& \cmarkempty	& \cmarkempty	& \cmarkempty	& \cmarkempty 	&  \cmarkempty 	& \cmarkempty 	& \cmarkempty 	& \cmarkempty 	& \cmarkempty 	& \multicolumn{1}{c:}{\circlet \phantom{*}} &  	& \cmarkempty & \cmarkempty & \cmarkempty \\
&			BTB & 			& \cmarkempty 	& \cmarkempty 	& \cmarkempty	& \cmarkempty & \multicolumn{1}{c:}{\circlet \phantom{*}} & & \cmarkempty	& \circletfill	 	& \cmarkempty 	&   \cmarkempty 	& \circletfill 	& \circletfill 	& \circletfill  	&  \cmarkempty & \cmarkempty & \multicolumn{1}{c:}{\circlet \phantom{*}} & & \cmarkempty	& \cmarkempty 	& \cmarkempty \\
&			BHB & 				& \cmarkempty 	& \cmarkempty 	& \cmarkempty & \cmarkempty & \multicolumn{1}{c:}{\circlet \phantom{*}} & 	& \cmarkempty 	& \cmarkempty 	& \cmarkempty 	& \cmarkempty 	& \cmarkempty	&  \cmarkempty 	&  \cmarkempty 	& \cmarkempty 	& \cmarkempty 	& \multicolumn{1}{c:}{\circlet \phantom{*}} &  	&  \cmarkempty & \cmarkempty & \cmarkempty \\
&			PHT & 				& \cmarkempty 	& \cmarkempty 	& \cmarkempty & \cmarkempty & \multicolumn{1}{c:}{\circlet \phantom{*}} & 	& \cmarkempty 	& \cmarkempty 	& \cmarkempty 	& \cmarkempty 	& \cmarkempty	&  \cmarkempty 	&  \cmarkempty 	& \cmarkempty 	& \cmarkempty 	& \multicolumn{1}{c:}{\circlet \phantom{*}} &  	&  \cmarkempty & \cmarkempty & \cmarkempty \\
&			RSB & 				& \cmarkempty 	& \cmarkempty 	& \cmarkempty & \cmarkempty & \multicolumn{1}{c:}{\circlet \phantom{*}} & 	& \circletfill 	& \circletfill 	 	& \cmarkempty 	&  \cmarkempty 	& \cmarkempty 	& \cmarkempty 	& \cmarkempty  	& \cmarkempty & \cmarkempty & \multicolumn{1}{c:}{\circlet \phantom{*}} & & \cmarkempty 	& \cmarkempty 	& \cmarkempty \\
&			AVX & 				& \cmarkempty 	& \cmarkempty 	& \cmarkempty & \cmarkempty & \multicolumn{1}{c:}{\circlet \phantom{*}} &	& \cmarkempty 	& \cmarkempty 	& \cmarkempty 	& \cmarkempty 	& \cmarkempty 	& \cmarkempty 	&  \cmarkempty 	& \cmarkempty 	& \cmarkempty 	& \multicolumn{1}{c:}{\circlet \phantom{*}} &  	& \cmarkempty & \cmarkempty & \cmarkempty \\
&			FPU & 				& \cmarkempty 	& \cmarkempty 	& \cmarkempty 	& \cmarkempty & \multicolumn{1}{c:}{\circlet \phantom{*}} & & \cmarkempty 	& \cmarkempty 	& \cmarkempty 	& \cmarkempty 	& \cmarkempty 	& \cmarkempty 	&  \cmarkempty 	& \cmarkempty 	& \cmarkempty 	& \multicolumn{1}{c:}{\circlet \phantom{*}} &  	& \cmarkempty & \cmarkempty & \cmarkempty \\
&			Ports & 				& \cmarkempty 	& \cmarkempty & \cmarkempty & \cmarkempty & \multicolumn{1}{c:}{\circlet \phantom{*}} & 	 	& \cmarkempty 	& \cmarkempty 	& \cmarkempty 	& \cmarkempty 	& \cmarkempty 	& \cmarkempty 	&  \cmarkempty 	& \cmarkempty  	& \cmarkempty	& \multicolumn{1}{c:}{\circlet \phantom{*}} &  	& \cmarkempty & \cmarkempty & \cmarkempty \\
&			\vspace{-12pt} & \multicolumn{6}{c:}{\,\raisebox{10pt}{\tikz{\draw [decorate,decoration={brace,amplitude=5pt},rotate=90] (0,0) -- (0,2.35);}}} & \multicolumn{11}{c:}{\raisebox{10pt}{\tikz{\draw [decorate,decoration={brace,amplitude=5pt},rotate=90] (0,0) -- (0,5);}}} & \multicolumn{4}{c}{\raisebox{10pt}{\tikz{\draw [decorate,decoration={brace,amplitude=5pt},rotate=90] (0,0) -- (0,1.35);}}\hspace{0.5pt}} \\
&			 & \multicolumn{6}{c:}{\hspace{0.75pt}\textbf{C1}} & \multicolumn{11}{c:}{\textbf{C2}} & \multicolumn{4}{c}{\hspace{0.5pt}\textbf{C3}} \\
			\bottomrule
		\end{tabular}
	}}
\end{center}
\vspace{-0.3cm}
  {\footnotesize
	A defense considers the microarchitectural element (\circletfill), partially considers it or same technique possible for it (\circletfillhl) or does not consider it at all (\circlet).
  }
    \vspace{-0.1cm}
\end{table}

\noindent
We categorize \Meltdown-type defenses into two categories:
\begin{enumerate}[nolistsep,align=left, leftmargin=13pt, labelwidth=8pt, itemindent=0pt]
  \item[\textbf{D1}:] Ensuring that architecturally inaccessible data remains inaccessible on the microarchitectural level.
  \item[\textbf{D2}:] Preventing the occurrence of faults.
\end{enumerate}

\subsection{Defenses for \Spectre}

\paragraph{C1: Mitigating or reducing accuracy of covert channels.}
Transient execution attacks use a covert channel to transfer a microarchitectural state change induced by the transient instruction sequence to the architectural level.
One approach in mitigating \Spectre-type attacks is reducing the accuracy of covert channels or preventing them.

\subparagraph{Hardware.}
One enabler of transient execution attacks is that the transient execution sequence introduces a microarchitectural state change the receiving end of the covert channel observes.
To secure CPUs, SafeSpec~\cite{khasawneh2018safespec} introduces shadow hardware structures used during transient execution.
Thereby, any microarchitectural state change can be squashed if the prediction of the CPU was incorrect.
While their prototype implementation protects only caches (and the TLB), other channels, \eg DRAM buffers~\cite{Pessl2016}, or execution unit congestion~\cite{Lipp2018meltdown,Aldaya2018,Bhattacharyya2019smother}, remain open.

Yan~\etal\cite{Yan2018InvisiSpec} proposed InvisiSpec, a method to make transient loads invisible in the cache hierarchy.
By using a \textit{speculative buffer}, all transiently executed loads are stored in this buffer instead of the cache.
Similar to SafeSpec, the buffer is invalidated if the prediction was incorrect.
However, if the prediction was correct, the content of the buffer is loaded into the cache.
For data coherency, InvisiSpec compares the loaded value during this process with the most recent, up-to-date value from the cache.
If a mismatch occurs, the transient load and all successive instructions are reverted.
Since InvisSpec only protects the caching hierarchy of the CPU, an attacker can still exploit other covert channels.

Kiriansky~\etal\cite{Kiriansky2018dawg} securely partition the cache across its ways.
With protection domains that isolate on a cache hit, cache miss and metadata level, cache-based covert channels are mitigated.
This does not only require changes to the cache and adaptions to the coherence protocol but also enforces the correct management of these domains in software.

Kocher~\etal\cite{Kocher2019spectre} proposed to limit data from entering covert channels through a variation of taint tracking.
The idea is that the CPU tracks data loaded during transient execution and prevents their use in subsequent operations.

\subparagraph{Software.}
Many covert channels require an accurate timer to distinguish microarchitectural states, \eg measuring the memory access latency to distinguish between a cache hit and cache miss.
With reduced timer accuracy an attacker cannot distinguish between microarchitectural states any longer, the receiver of the covert channel cannot deduce the sent information.
To mitigate browser-based attacks, many web browsers reduced the accuracy of timers in JavaScript by adding jitter~\cite{Microsoft2018edge,Pizlo18,Chromium18mitigations,Wagner18firefox}.
However, Schwarz~\etal\cite{Schwarz2017Timers} demonstrated that timers can be constructed in many different ways and, thus, further mitigations are required~\cite{Schwarz2018JSZ}.
While Chrome initially disabled \texttt{SharedArrayBuffers} in response to Meltdown and Spectre~\cite{Chromium18mitigations}, this timer source has been re-enabled with the introduction of site-isolation~\cite{Chromium18enablebuffers}.

\NetSpectre requires different strategies due to its remote nature.
Schwarz~\etal\cite{Schwarz2018netspectre} propose to detect the attack using DDoS detection mechanisms or adding noise to the network latency.
By adding noise, an attacker needs to record more traces.
Adding enough noise makes the attack infeasible in practice as the amount of traces as well as the time required for averaging it out becomes too large~\cite{Varda2018cloudflare}.

\paragraph{C2: Mitigating or aborting speculation if data is potentially accessible during transient execution.}

Since \Spectre-type attacks exploit different prediction mechanisms used for speculative execution, an effective approach would be to disable speculative execution entirely~\cite{Kocher2019spectre, Suse2018DisableBP}.
As the loss of performance for commodity computers and servers would be too drastic, another proposal is to disable speculation only while processing secret data.

\subparagraph{Hardware.}
A building blocks for some variants of \Spectre is branch poisoning (an attacker mistrains a prediction mechanism, \cf~\cref{sec:spectre}).
To deal with mistraining, both Intel and AMD extended the instruction set architecture (ISA) with a mechanism for controlling indirect branches~\cite{AMDSpecAnalysis,IntelMitigations}.
The proposed addition to the ISA consists of three controls:

\begin{compactitem}
\item Indirect Branch Restricted Speculation (IBRS) prevents indirect branches executed in privileged code from being influenced by those in less privileged code.
To enforce this, the CPU enters the IBRS mode which cannot be influenced by any operations outside of it.
\item Single Thread Indirect Branch Prediction (STIBP) restricts sharing of branch prediction mechanisms among code executing across hyperthreads.
\item The Indirect Branch Predictor Barrier (IBPB) prevents code that executes before it from affecting the prediction of code following it by flushing the BTB.
\end{compactitem}

For existing ARM implementations, there are no generic mitigation techniques available.
However, some CPUs implement specific controls that allow invalidating the branch predictor which should be used during context switches~\cite{ARMSpecAnalysis_whitepaper}.
On Linux, those mechanisms are enabled by default~\cite{ARMSpectreLinuxv2Hardening}.
With the ARMv8.5-A instruction set~\cite{arm_arch_manualv85}, ARM introduces a new barrier (\inst{sb}) to limit speculative execution on following instructions.
Furthermore, new system registers allow to restrict speculative execution and new prediction control instructions prevent control flow predictions (\inst{cfp}), data value prediction (\inst{dvp}) or cache prefetch prediction (\inst{cpp})~\cite{arm_arch_manualv85}.

To mitigate \SpectreSTL, ARM introduced a new barrier called \inst{SSBB} that prevents a load following the barrier from bypassing a store using the same virtual address before it~\cite{ARMSpecAnalysis_whitepaper}.
For upcoming CPUs, ARM introduced Speculative Store Bypass Safe (SSBS); a configuration control register to prevent the re-ordering of loads and stores~\cite{ARMSpecAnalysis_whitepaper}.
Likewise, Intel~\cite{IntelMitigations} and AMD~\cite{AMDssbd_whitepaper}
provide Speculative Store Bypass Disable (SSBD) microcode updates that mitigate \SpectreSTL.

As an academic contribution, plausible hardware mitigations have furthermore been proposed~\cite{Kiriansky2018speculative}
to prevent transient computations on out-of-bounds writes (\SpectrePHT).

\subparagraph{Software.}
Intel and AMD proposed to use serializing instructions like \inst{lfence} on both outcomes of a branch~\cite{AMDSpecAnalysis,Intel2018white_paper}.
ARM introduced a full data synchronization barrier (\inst{DSB SY}) and an instruction synchronization barrier (ISB) that can be used to prevent speculation~\cite{ARMSpecAnalysis_whitepaper}.
Unfortunately, serializing every branch would amount to completely disabling branch prediction, severely reducing performance~\cite{Intel2018white_paper}.
Hence, Intel further proposed to use static analysis~\cite{Intel2018white_paper} to minimize the number of serializing instructions introduced.
Microsoft uses the static analyzer of their C Compiler MSVC~\cite{Microsoft2018Spectre} to detect known-bad code patterns and insert \inst{lfence} instructions automatically.
Open Source Security Inc.~\cite{Grsecurity2018Respectre} use a similar approach using static analysis.
Kocher~\cite{Kocher2018mitigations} showed that this approach misses many gadgets that can be exploited.

Serializing instructions can also reduce the effect of indirect branch poisoning.
By inserting it before the branch, the pipeline prior to it is cleared, and the branch is resolved quickly~\cite{AMDSpecAnalysis}.
This, in turn, reduces the size of the speculation window in case that misspeculation occurs.

While \inst{lfence} instructions stop speculative execution, Schwarz~\etal~\cite{Schwarz2018netspectre} showed they do not stop microarchitectural behaviors happening before execution.
This, for instance, includes powering up the AVX functional units, instruction cache fills, and iTLB fills which still leak data.

Evtyushkin~\etal\cite{Evtyushkin2018BranchScope} propose a similar method to serializing instructions, where a developer annotates potentially leaking branches.
When indicated, the CPU should not predict the outcome of these branches and thus stop speculation.

Additionally to the serializing instructions, ARM also introduced a new barrier (\inst{CSDB}) that in combination with conditional selects or moves controls speculative execution~\cite{ARMSpecAnalysis_whitepaper}.

Speculative Load Hardening (SLH) is an approach used by LLVM and was proposed by Carruth~\cite{Carruth2018Hardening}.
Using this idea, loads are checked using branchless code to ensure that they are executing along a valid control flow path.
To do this, they transform the code at the compiler level and introduce a data dependency on the condition.
In the case of misspeculation, the pointer is zeroed out, preventing it from leaking data through speculative execution.
One prerequisite for this approach is hardware that allows the implementation of a branchless and unpredicted conditional update of a register's value.
As of now, the feature is only available in LLVM for x86 as the patch for ARM is still under review.
GCC adopted the idea of SLH for their implementation, supporting both x86 and ARM.
They provide a builtin function to either emit a speculation barrier or return a safe value if it determines that the instruction is transient~\cite{gcc2018speculation}.

Oleksenko~\etal\cite{Oleksenko2018ysnb} propose an approach similar to Carruth~\cite{Carruth2018Hardening}.
They exploit that CPUs have a mechanism to detect data dependencies between instructions and introduce such a dependency on the comparison arguments.
This ensures that the load only starts when the comparison is either in registers or the L1 cache, reducing the speculation window to a non-exploitable size.
They already note that their approach is highly dependent on the ordering of instructions as the CPU might perform the load before the comparison.
In that case, the attack would still be possible.

Google proposes a method called \textit{retpoline}~\cite{Turner2018retpoline}, a code sequence that replaces indirect branches with return instructions, to prevent branch poisoning.
This method ensures that return instructions always speculate into an endless loop through the RSB.
The actual target destination is pushed on the stack and returned to using the \inst{ret} instruction.
For retpoline, Intel~\cite{Intel2018retpoline} notes that in future CPUs that have Control-flow Enforcement Technology (CET) capabilities to defend against ROP attacks, retpoline might trigger false positives in the CET defenses.
To mitigate this possibility, future CPUs also implement hardware defenses for \SpectreBTB called \textit{enhanced IBRS}~\cite{Intel2018retpoline}.

On Skylake and newer architectures, Intel~\cite{Intel2018retpoline} proposes RSB stuffing to prevent an RSB underfill and the ensuing fallback to the BTB.
Hence, on every context switch into the kernel, the RSB is filled with the address of a benign gadget.
This behavior is similar to retpoline.
For Broadwell and older architectures, Intel~\cite{Intel2018retpoline} provided a microcode update to make the \inst{ret} instruction predictable, enabling retpoline to be a robust defense against \SpectreBTB.
Windows has also enabled retpoline on their systems~\cite{Microsoft2019retpoline}.

\paragraph{C3: Ensuring that secret data cannot be reached.}
Different projects use different techniques to mitigate the problem of \Spectre.
WebKit employs two such techniques to limit the access to secret data~\cite{Pizlo18}.
WebKit first replaces array bound checks with index masking.
By applying a bit mask, WebKit cannot ensure that the access is always in bounds, but introduces a maximum range for the out-of-bounds violation.
In the second strategy, WebKit uses a pseudo-random \textit{poison value} to protect pointers from misuse.
Using this approach, an attacker would first have to learn the poison value before he can use it.
The more significant impact of this approach is that mispredictions on the branch instruction used for type checks results in the wrong type being used for the pointer.

Google proposes another defense called \textit{site isolation}~\cite{Chromium2018SiteIsolation}, which is now enabled in Chrome by default.
Site isolation executes each site in its own process and therefore limits the amount of data that is exposed to side-channel attacks.
Even in the case where the attacker has arbitrary memory reads, he can only read data from its own process.

Kiriansky and Waldspurger~\cite{Kiriansky2018speculative} propose to restrict access to sensitive data by using protection keys like Intel Memory Protection Key (MPK) technology~\cite{Intel_vol3}.
They note that by using \SpectrePHT an attacker can first disable the protection before reading the data.
To prevent this, they propose to include an \inst{lfence} instruction in \inst{wrpkru}, an instruction used to modify protection keys.

\subsection{Defenses for \Meltdown}
\paragraph{D1: Ensuring that architecturally inaccessible data remains inaccessible on the microarchitectural level.}

The fundamental problem of \Meltdown-type attacks is that the CPU allows the transient instruction stream to compute on architecturally inaccessible values, and hence, leak them.
By assuring that execution does not continue with unauthorized data after a fault, such attacks can be mitigated directly in silicon.
This design is enforced in AMD processors~\cite{AMDSpecAnalysis}, and more recently also in Intel processors from Whiskey Lake onwards that enumerate \rdclno{} support~\cite{IntelMitigations}.
However, mitigations for existing microarchitectures are necessary, either through microcode updates, or operating-system-level software workarounds.
These approaches aim to keep architecturally inaccessible data also inaccessible at the microarchitectural level.

Gruss~\etal originally proposed KAISER~\cite{Gruss2017KASLR,Gruss2016Prefetch} to mitigate side-channel attacks defeating KASLR.
However, it also defends against \MeltdownUS attacks by preventing kernel secrets from being mapped in user space.
Besides its performance impact, KAISER has one practical limitation~\cite{Lipp2018meltdown,Gruss2017KASLR}.
For x86, some privileged memory locations must always be mapped in user space.
KAISER is implemented in Linux as kernel page-table isolation (KPTI)~\cite{LWN_kpti} and has also been backported to older versions.
Microsoft provides a similar patch as of Windows 10 Build 17035~\cite{Ionescu2017Twitter} and Mac OS X and iOS have similar patches~\cite{Ionescu2017TwitterApple}.

For \MeltdownGP, where the attacker leaks the contents of system registers that are architecturally not accessible in its current privilege level, Intel released microcode updates~\cite{Intel2018white_paper}.
While AMD is not susceptible~\cite{AMD2018_variant_3a}, ARM incorporated mitigations in future CPU designs and suggests to substitute the register values with dummy values on context switches for CPUs where mitigations are not available~\cite{ARMSpecAnalysis_whitepaper}.

Preventing the access-control race condition exploited by \Foreshadow and \Meltdown may not be feasible with microcode updates~\cite{Vanbulck2018}.
Thus, Intel proposes a multi-stage approach to mitigate \Foreshadow (L1TF) attacks on current CPUs~\cite{Intel2018l1tf,Weisse2018foreshadowNG}.
First, to maintain process isolation, the operating system has to sanitize the physical address field of unmapped page-table entries.
The kernel either clears the physical address field, or sets it to non-existent physical memory.
In the case of the former, Intel suggests placing 4 KB of dummy data at the start of the physical address space, and clearing the PS bit in page tables to prevent attackers from exploiting huge pages.

For SGX enclaves or hypervisors, which cannot trust the address translation performed by an untrusted OS, Intel proposes to either store secrets in uncacheable memory (as specified in the PAT or the MTRRs), or flush the L1 data cache when switching protection domains.
With recent microcode updates, L1 is automatically flushed upon enclave exit, and hypervisors can additionally flush L1 before handing over control to an untrusted virtual machine.
Flushing the cache is also done upon exiting System Management Mode (SMM) to mitigate \ForeshadowNG attacks on SMM.

To mitigate attacks across logical cores, Intel supplied a microcode update to ensure that different SGX attestation keys are derived when hyperthreading is enabled or disabled.
To ensure that no non-SMM software runs while data belonging to SMM are in the L1 data cache, SMM software must rendezvous all logical cores upon entry and exit.
According to Intel, this is expected to be the default behavior for most SMM software~\cite{Intel2018l1tf}.
To protect against \ForeshadowNG attacks when hyperthreading is enabled, the hypervisor must ensure that no hypervisor thread runs on a sibling core with an untrusted VM.

\paragraph{D2: Preventing the occurrence of faults.}\label{sec:meltdown-defense-d2}
Since \Meltdown-type attacks exploit delayed exception handling in the CPU,
another mitigation approach is to prevent the occurrence of a fault in the first place.
Thus, accesses which would normally fault, become (both architecturally and microarchitecturally) valid accesses but do not leak secret data.

One example of such behavior are SGX's abort page semantics, where accessing enclave memory from the outside returns -1 instead of faulting.
Thus, SGX has inadvertent protection against \MeltdownUS.
However, the \Foreshadow~\cite{Vanbulck2018} attack showed that it is possible to actively provoke another fault by unmapping the enclave page, making SGX enclaves susceptible to the \MeltdownP variant.

Preventing the fault is also the countermeasure for \MeltdownNM~\cite{Stecklina2018LazyFP} that is deployed since Linux 4.6~\cite{lklm18eagerswitching}.
By replacing lazy switching with eager switching, the FPU is always available, and access to the FPU can never fault.
Here, the countermeasure is effective, as there is no other way to provoke a fault when accessing the FPU.


\subsection{Evaluation of Defenses}
\label{sec:evaluation}
\paragraph{\Spectre Defenses.}
We evaluate defenses based on their capabilities of mitigating \Spectre attacks.
Defenses that require hardware modifications are only evaluated theoretically.
In addition, we discuss which vendors have CPUs vulnerable to what type of \Spectre- and \Meltdown-type attack.
The results of our evaluation are shown in \cref{tab:spectre_defenses_vulnerability_assessment}.

Several defenses only consider a specific covert channel (see \cref{tab:spectre_defenses_target_me_element}), \ie they only try to prevent an attacker from recovering the data using a specific covert channel instead of targeting the root cause of the vulnerability.
Therefore, they can be subverted by using a different one.
As such, they can not be considered a reliable defense.
Other defenses only limit the amount of data that can be leaked~\cite{Pizlo18,Chromium2018SiteIsolation} or simply require more repetitions on the attacker side~\cite{Schwarz2018netspectre,Varda2018cloudflare}.
Therefore, they are only partial solutions.
RSB stuffing only protects a cross-process attack but does not mitigate a same-process attack.
Many of the defenses are not enabled by default or depend on the underlying hardware and operating system~\cite{AMDSpecAnalysis,IntelMitigations,AMDssbd_whitepaper,ARMSpecAnalysis_whitepaper}.
With serializing instructions~\cite{AMDSpecAnalysis,Intel2018white_paper,ARMSpecAnalysis_whitepaper} after a bounds check, we were still able to leak data on Intel and ARM (only with \inst{DSB SY+ISH} instruction) through a single memory access and the TLB.
On ARM, we observed no leakage following a \inst{CSDB} barrier in combination with conditional selects or moves.
We also observed no leakage with SLH, although the possibility remains that our experiment failed to bypass the mitigation.
Taint tracking theoretically mitigates all forms of \Spectre-type attacks as data that has been tainted cannot be used in a transient execution.
Therefore, the data does not enter a covert channel and can subsequently not be leaked.

\paragraph{\Meltdown Defenses.}
We verified whether we can still execute \Meltdown-type attacks on a fully-patched system.
On a Ryzen Threadripper 1920X, we were still able to execute Meltdown-BND.
On an i5-6200U (Skylake), an i7-8700K (Coffee Lake), and an i7-8565U (Whiskey Lake), we were able to successfully run a Meltdown-MPX, Meltdown-BND, and \MeltdownRW attack.
Additionally to those, we were also able to run a \MeltdownPK attack on an Amazon EC2 C5 instance (Skylake-SP).
Our results indicate that current mitigations only prevent \Meltdown-type attacks that cross the current privilege level.
We also tested whether we can still successfully execute a \MeltdownUS attack on a recent Intel Whiskey Lake CPU without KPTI enabled, as Intel claims these processors are no longer vulnerable.
In our experiments, we were indeed not able to leak any data on such CPUs but encourage other researchers to further investigate newer processor generations.

\begin{table}[t]
	\setlength{\aboverulesep}{0pt}
	\setlength{\belowrulesep}{0pt}
	\caption{Spectre defenses and which attacks they mitigate.}\label{tab:spectre_defenses_vulnerability_assessment}
\vspace{-0.2cm}
\adjustbox{max width=0.96\hsize}{\small{
		\setlength\tabcolsep{1.5pt}
		\begin{tabular}{cc llllllllllllllllll}
      & \diagbox{\textbf{Attack}}{\textbf{Defense}} & \rotatebox{70}{\hspace{-0.66em}\texttt{InvisiSpec}} \hspace{-20em} & \rotatebox{70}{\hspace{-0.66em}\texttt{SafeSpec}} \hspace{-20em} & \rotatebox{70}{\hspace{-0.66em}\texttt{DAWG}} \hspace{-20em} & \rotatebox{70}{\hspace{-0.66em}\textit{RSB Stuffing}} \hspace{-20em} & \rotatebox{70}{\hspace{-0.66em}\textit{Retpoline}} \hspace{-20em} & \rotatebox{70}{\hspace{-0.66em}\textit{Poison Value}} \hspace{-20em} & \rotatebox{70}{\hspace{-0.66em}\textit{Index Masking}} \hspace{-20em} & \rotatebox{70}{\hspace{-0.66em}\textit{Site Isolation}} \hspace{-20em} & \rotatebox{70}{\hspace{-0.66em}\textit{SLH}} \hspace{-20em} & \rotatebox{70}{\hspace{-0.66em}\texttt{YSNB}} \hspace{-20em} & \rotatebox{70}{\hspace{-0.66em}\textit{IBRS}} \hspace{-20em} & \rotatebox{70}{\hspace{-0.66em}\textit{STIBP}} \hspace{-20em} & \rotatebox{70}{\hspace{-0.66em}\textit{IBPB}} \hspace{-20em} & \rotatebox{70}{\hspace{-0.66em}\textit{Serialization}} \hspace{-20em} & \rotatebox{70}{\hspace{-0.66em}\texttt{Taint Tracking}} \hspace{-20em} & \rotatebox{70}{\hspace{-0.66em}\textit{Timer Reduction}} \hspace{-20em} & \rotatebox{70}{\hspace{-0.66em}\texttt{Sloth}} \hspace{-20em} & \rotatebox{70}{\hspace{-0.66em}\textit{SSBD/SSBB}} \hspace{-20em} \\
			\toprule \vspace{-0.3cm}\\
			\multirow{4}{*}{Intel} & \SpectrePHT	& \squad 	& \squad	& \squad 	& \rhombus 	& \rhombus 	& \circletfill & \circletfillhl & \circletfillhl & \circletfill & \circlet & \rhombus & \rhombus & \rhombus &  \circletfillhl	& \squadfill 	& \circletfillhl 	& \squadfillhl 	&  \rhombus \\
			& \SpectreBTB	& \squad 	& \squad	& \squad 	& \rhombus 	& \circletfill 	& \rhombus 	& \rhombus 	& \circletfillhl 	& \rhombus 	& \rhombus 	& \circletfill 	& \circletfillhl 	& \circletfillhl 	&  \rhombus	& \squadfill 	& \circletfillhl 	& \rhombus 	&  \rhombus \\
			& \SpectreRSB	& \squad 	& \squad	& \squad 	& \circletfillhl 	& \rhombus 	& \rhombus 	& \rhombus 	& \circletfillhl 	& \rhombus 	& \rhombus 	& \rhombus 	& \rhombus 	& \rhombus 	&  \rhombus	& \squadfill 	& \circletfillhl 	& \rhombus 	&  \rhombus \\
			& \SpectreSTL	& \squad 	& \squad	& \squad 	& \rhombus 	& \rhombus 	& \rhombus 	& \rhombus 	& \circletfillhl 	& \rhombus 	& \rhombus	& \rhombus 	& \rhombus 	& \rhombus 	&  \rhombus & \squadfill 	& \circletfillhl 	& \squadfill 	&  \circletfill	\\
			\midrule
			\multirow{4}{*}{ARM} & \SpectrePHT	& \squad 	& \squad	& \squad 	& \rhombus 	& \rhombus 	& \circletfill 	& \circletfillhl 	& \circletfillhl 	& \circletfill 	& \circlet 	& \rhombus 	& \rhombus 	&  \rhombus	&  \circletfillhl & \squadfill 	& \circletfillhl 	& \squadfillhl 	&  \rhombus	\\
			& \SpectreBTB	& \squad 	& \squad	& \squad 	& \rhombus 	& \circletfill 	& \rhombus 	& \rhombus	& \circletfillhl 	& \rhombus 	& \rhombus 	& \rhombus	& \rhombus 	& \rhombus 	&  \rhombus & \squadfill 	& \circletfillhl 	& \rhombus 	&  \rhombus	\\
			& \SpectreRSB	& \squad 	& \squad	& \squad 	& \circletfillhl 	& \rhombus 	& \rhombus 	& \rhombus 	& \circletfillhl 	& \rhombus 	& \rhombus 	& \rhombus 	& \rhombus 	& \rhombus 	&  \rhombus & \squadfill 	& \circletfillhl 	& \rhombus 	&  \rhombus	\\
			& \SpectreSTL	& \squad 	& \squad	& \squad 	& \rhombus 	& \rhombus 	& \rhombus 	& \rhombus 	& \circletfillhl 	& \rhombus 	& \rhombus 	& \rhombus 	& \rhombus 	& \rhombus 	&  \rhombus	& \squadfill 	& \circletfillhl 	& \squadfill 	&  \circletfill \\
			\midrule
			\multirow{4}{*}{AMD} & \SpectrePHT	& \squad 	& \squad	& \squad 	& \rhombus 	& \rhombus 	& \circletfill 	& \circletfillhl 	& \circletfillhl 	& \circletfill 	& \circlet 	& \rhombus 	& \rhombus 	& \rhombus 	&  \circletfillhl	& \squadfill 	& \circletfillhl 	& \squadfillhl 	&  \rhombus \\
			& \SpectreBTB	& \squad 	& \squad	& \squad 	& \rhombus 	& \circletfill 	& \rhombus 	& \rhombus 	& \circletfillhl 	& \rhombus 	& \rhombus 	& \squadfill & \squadfillhl &  \squadfillhl	&  \rhombus & \squadfill 	& \circletfillhl 	& \rhombus 	&  \rhombus	\\
			& \SpectreRSB	& \squad 	& \squad	& \squad 	& \circletfillhl 	& \rhombus 	& \rhombus 	& \rhombus 	& \circletfillhl 	& \rhombus	& \rhombus 	& \rhombus 	& \rhombus 	& \squadfillhl 	&  \rhombus	 & \squadfill 	& \circletfillhl 	& \rhombus 	&  \rhombus \\
			& \SpectreSTL	& \squad 	& \squad	& \squad 	& \rhombus 	& \rhombus 	& \rhombus 	& \rhombus 	& \circletfillhl 	& \rhombus 	& \rhombus 	& \rhombus	& \rhombus 	& \rhombus 	&  \rhombus	 & \squadfill 	& \circletfillhl 	& \squadfill 	&  \circletfill \\
			\bottomrule
		\end{tabular}
	}}
\vspace{0.05cm}\\
  {\footnotesize
Symbols show if an attack is mitigated (\circletfill), partially mitigated (\circletfillhl), not mitigated (\circlet), theoretically mitigated (\squadfill), theoretically impeded (\squadfillhl), not theoretically impeded (\squad), or out of scope (\rhombus).
Defenses in \textit{italics} are production-ready, while \texttt{typeset} defenses are academic proposals.
  }
    \vspace{-0.1cm}
\end{table}

\subsection{Performance Impact of Countermeasures}
There have been several reports on performance impacts of selected countermeasures.
Some report the performance impact based on real-world scenarios (top of \cref{tab:performance-countermeasures}) while others use a specific benchmark that might not resemble real-world usage (lower part of \cref{tab:performance-countermeasures}).
Based on the different testing scenarios, the results are hard to compare.
To further complicate matters, some countermeasures require hardware modifications that are not available, and it is therefore hard to verify the performance loss.

One countermeasure that stands out with a huge decrease in performance is serialization and highlights the importance of speculative execution to improve CPU performance.
Another interesting countermeasure is KPTI.
While it was initially reported to have a huge impact on performance, recent work shows that the decrease is almost negligible on systems that support PCID~\cite{Gregg2018kpti}.
To mitigate Spectre and Meltdown, current systems rely on a combination of countermeasures.
To show the overall decrease on a Linux 4.19 kernel with the default mitigations enabled, Larabel~\cite{Larabel2018cost} performed multiple benchmarks to determine the impact.
On Intel, the slowdown was 7-16\% compared to a non-mitigated kernel, on AMD it was 3-4\%.

Naturally, the question arises which countermeasures to enable.
For most users, the risk of exploitation is low, and default software mitigations as provided by Linux, Microsoft, or Apple likely are sufficient.
This is likely the optimum between potential attacks and reasonable performance.
For data centers, it is harder as it depends on the needs of their customers and one has to evaluate this on an individual basis.

\begin{table}[t]
	\setlength{\aboverulesep}{0pt}
	\setlength{\belowrulesep}{0pt}
	\caption{Reported performance impacts of countermeasures.
  Top shows performance impact in real-world scenarios while the bottom shows it on a specific benchmark.}\label{tab:performance-countermeasures}
\begin{center}
\vspace{-0.4cm}
\adjustbox{max width=1\hsize}{\small{
		\setlength\tabcolsep{3pt}
		\begin{tabular}{llll}
            \textbf{Defense Evaluation} & \textbf{Penalty} & \textbf{Benchmark} \\
			\toprule \vspace{-0.3cm}\\
      KAISER/KPTI~\cite{Gruss2018Kernel} & \SIx{0}--\SI{2.6}{\percent} & System call rates\\
      Retpoline~\cite{Carruth2018retpoline_patch}  & \SIx{5}--\SI{10}{\percent} & Real-world workload servers \\
      Site Isolation~\cite{Chromium2018SiteIsolation} & \SIx{10}--\SI{13}{\percent} & Memory overhead\\
      \midrule
      InvisiSpec~\cite{Yan2018InvisiSpec}  & \SI{22}{\percent} & SPEC \\
      SafeSpec~\cite{khasawneh2018safespec}& -\SI{3}{\percent} & SPEC on MARSSx86 \\
      DAWG~\cite{Kiriansky2018dawg}        & \SIx{1}--\SI{15}{\percent} & PARSEC~, GAPBS \\
      SLH~\cite{Carruth2018Hardening} & \SIx{29}--\SI{36.4}{\percent}& Google microbenchmark suite \\
      YSNB~\cite{Oleksenko2018ysnb} & \SI{60}{\percent} & Phoenix \\
      IBRS~\cite{Tkachenko2018ibrs_performance} & \SIx{20}--\SI{30}{\percent} & Sysbench 1.0.11 \\
      STIBP~\cite{Larabel2018stibp} & \SIx{30}--\SI{50}{\percent} & Rodinia OpenMP, DaCapo\\
      Serialization~\cite{Carruth2018Hardening} & \SIx{62}--\SI{74.8}{\percent} & Google microbenchmark suite \\
      SSBD/SSBB~\cite{Culbertson2018performance} & \SIx{2}--\SI{8}{\percent} & SYSmark 2018, SPEC integer \\
      L1TF Mitigations~\cite{Intel2018L1TFperf} & -\SIx{3}--\SI{31}{\percent} & SPEC \\
			\bottomrule
		\end{tabular}
	}}
	\end{center}
	\vspace{-0.7cm}
\end{table}

\section{Future Work and Conclusion}
\label{sec:future-work-conclusion}
\paragraph{Future Work.}
For Meltdown-type attacks, it is important to determine where data is actually leaked from.
For instance, Lipp~\etal\cite{Lipp2018meltdown} demonstrated that \MeltdownUS can not only leak data from the L1 data cache and main memory but even from memory locations that are explicitly marked as \enquote{uncacheable} and are hence served from the Line Fill Buffer (LFB).
\footnote{The initial \MeltdownUS disclosure (December 2017) and subsequent paper~\cite{Lipp2018meltdown} already made clear that Meltdown-type leakage is \emph{not} limited to the L1 data cache.
We sent Intel a PoC leaking uncacheable-typed memory locations from a concurrent hyperthread on March 28, 2018.
We clarified to Intel on May 30, 2018, that we attribute the source of this leakage to the LFB.
In our experiments, this works identically for \MeltdownP (Foreshadow).
This issue was acknowledged by Intel, tracked under CVE-2019-11091, and remained under embargo until May 14, 2019.}
In future work, other Meltdown-type attacks should be tested to determine whether they can also leak data from different microarchitectural buffers.
In this paper, we presented a small evaluation of the prevalence of gadgets in real-world software.
Future work should develop methods for automating the detection of gadgets and extend the analysis on a larger amount of real-world software.
We have also discussed mitigations and shown that some of them can be bypassed or do not target the root cause of the problem.
We encourage both offensive and defensive research that may use our taxonomy as a guiding principle to discover new attack variants, and develop mitigations that target the root cause of transient information leakage.

\paragraph{Conclusion.}
Transient instructions reflect unauthorized computations out of the program's intended code and/or data paths.
We presented a systematization of transient execution attacks.
Our systematization uncovered 6 (new) transient execution attacks (Spectre and Meltdown variants) which have been overlooked and have not been investigated so far.
We demonstrated these variants in practical proof-of-concept attacks and evaluated their applicability to Intel, AMD, and ARM CPUs.
We also presented a short analysis and classification of gadgets as well as their prevalence in real-world software.
We also systematically evaluated defenses, discovering that some transient execution attacks are not successfully mitigated by the rolled out patches and others are not mitigated because they have been overlooked.
Hence, we need to think about future defenses carefully and plan to mitigate attacks and variants that are yet unknown.

\section*{Acknowledgments}
We want to thank the anonymous reviewers and especially our shepherd, Jonathan McCune, for their helpful comments and suggestions that substantially helped in improving the paper.
This work has been supported by the Austrian Research Promotion Agency (FFG) via the K-project DeSSnet, which is funded in the context of COMET – Competence Centers for Excellent Technologies by BMVIT, BMWFW, Styria and Carinthia.
This work has been supported by the Austrian Research Promotion Agency (FFG) via the project ESPRESSO, which is funded by the province of Styria and the Business Promotion Agencies of Styria and Carinthia.
This project has received funding from the European Research Council (ERC) under the European Union's Horizon 2020 research and innovation programme (grant agreement No 681402).
This research received funding from the Research Fund KU Leuven, and Jo Van Bulck is supported by the Research Foundation~--~Flanders~(FWO).
Evtyushkin acknowledges the start-up grant from the College of William and Mary.
Additional funding was provided by generous gifts from ARM and Intel.
Any opinions, findings, and conclusions or recommendations expressed in this paper are those of the authors and do not necessarily reflect the views of the funding parties.

{\footnotesize \fontsize{7}{8}\selectfont \bibliographystyle{acm-url}
\bibliography{main}}

\end{document}

%% file: images/classification_small.tikz
\begin{tikzpicture}[node distance=0.75cm,transform shape,scale=0.7]
\tikzstyle{small}  = [rectangle, rounded corners, minimum width=2.5cm, minimum height=.6cm,text centered, draw=black, fill=white]
\tikzstyle{large}  = [rectangle, rounded corners, minimum width=3.5cm, minimum height=.6cm,text centered, draw=black, fill=white]
\tikzstyle{largelow}  = [font=\scriptsize,rectangle, rounded corners, minimum width=3.5cm, minimum height=.1cm,text centered, draw=black, fill=white,node distance=0.55cm]
\tikzstyle{works}  = [fill=red!30!white,thick,font=\bfseries]
\tikzstyle{workslow}  = [fill=red!30!white,thick,font=\scriptsize\bfseries]
\tikzstyle{fails}  = [densely dashed,pattern=north west lines,pattern color=green!60!white]
\tikzstyle{group}  = [fill=blue!20!white]
\tikzstyle{arrow}  = [thick,->,>=stealth,in=180,out=0,looseness=0.6]
\tikzstyle{arrow_regular}  = [thick,->,>=stealth,in=180,out=0]
\tikzstyle{decide} = [font=\em\color{darkgray}]

\usetikzlibrary{shapes.geometric, arrows, patterns}

\node (root) [small, minimum width=1.6cm] {\parbox{1.6cm}{\centering\textbf{Transient\\cause?}}};
\node (spectretype) [small,group,right of=root,yshift=2.5cm,xshift=1cm] {Spectre-type};
\node [decide,below of=spectretype,xshift=0.75cm,yshift=1.75cm] {\parbox{2.5cm}{microarchitec-tural buffer}};

\node (meltdowntype) [small,group,below of=spectretype,yshift=-4cm] {Meltdown-type};
\node [decide,above of=meltdowntype,xshift=1cm,yshift=0.5cm] {fault type};

\node (spectrepht) [large,group,right of=spectretype,xshift=3.5cm,yshift=2cm] {Spectre-PHT};
\node (spectrebtb) [large,group,below of=spectrepht] {Spectre-BTB};
\node (spectrersb) [large,group,below of=spectrebtb] {Spectre-RSB};
\node (spectrestl) [large,works,below of=spectrersb] {Spectre-STL~\cite{Horn2018spectre4}};

\node [decide,above of=spectrepht,xshift=1.85cm,yshift=0.2cm] {\parbox{2.5cm}{mistraining\\ strategy}};

\newcommand{\cas}{Cross-address-space}
\newcommand{\sas}{Same-address-space}

\node (spectrephtca) [large,group,right of=spectrepht,xshift=3.75cm,yshift=1cm] {\cas{}};
\node (spectrephtsa) [large,group,below of=spectrephtca] {\sas{}};
\node (spectrephtcaip) [large,works,right of=spectrephtsa,xshift=3.5cm,yshift=1.5cm]
    {PHT-CA-IP $\medstar$};
\node (spectrephtcaop) [large,works,below of=spectrephtcaip] {PHT-CA-OP $\medstar$};
\node (spectrephtsaip) [large,works,below of=spectrephtcaop] {PHT-SA-IP~\cite{Kocher2019spectre,Kiriansky2018speculative}};
\node (spectrephtsaop) [large,works,below of=spectrephtsaip] {PHT-SA-OP $\medstar$};

\node [decide,above of=spectrephtca,xshift=-.55cm] {in-place (IP) vs., out-of-place (OP)};

\node (spectrebtbca) [large,group,right of=spectrebtb,xshift=3.75cm,yshift=-0.75cm] {\cas{}};
\node (spectrebtbsa) [large,group,below of=spectrebtbca] {\sas{}};
\node (spectrebtbcaip) [large,works,below of=spectrephtsaop] {BTB-CA-IP~\cite{Kocher2019spectre,Chen2018SGXpectre}};
\node (spectrebtbcaop) [large,works,below of=spectrebtbcaip] {BTB-CA-OP~\cite{Kocher2019spectre}};
\node (spectrebtbsaip) [large,works,below of=spectrebtbcaop] {BTB-SA-IP $\medstar$};
\node (spectrebtbsaop) [large,works,below of=spectrebtbsaip] {BTB-SA-OP \cite{Chen2018SGXpectre}};

\node (spectrersbca) [large,group,right of=spectrersb,xshift=3.75cm,yshift=-2.1cm] {\cas{}};
\node (spectrersbsa) [large,group,below of=spectrersbca] {\sas{}};
\node (spectrersbcaip) [large,works,below of=spectrebtbsaop] {RSB-CA-IP~\cite{Maisuradze2018spectre5,Koruyeh2018spectre5}};
\node (spectrersbcaop) [large,works,below of=spectrersbcaip] {RSB-CA-OP~\cite{Koruyeh2018spectre5}};
\node (spectrersbsaip) [large,works,below of=spectrersbcaop] {RSB-SA-IP~\cite{Maisuradze2018spectre5}};
\node (spectrersbsaop) [large,works,below of=spectrersbsaip] {RSB-SA-OP~\cite{Maisuradze2018spectre5,Koruyeh2018spectre5}};

\node (meltdownnm) [large,works,right of=meltdowntype,xshift=3.5cm,yshift=1.5cm] {Meltdown-NM~\cite{Stecklina2018LazyFP}};
\node (meltdownac) [large,fails,below of=meltdownnm] {Meltdown-AC $\medwhitestar$};
\node (meltdownde) [large,fails,below of=meltdownac] {Meltdown-DE $\medwhitestar$};
\node (meltdownpf) [large,group,below of=meltdownde] {Meltdown-PF};
\node (meltdownud) [large,fails,below of=meltdownpf] {Meltdown-UD $\medwhitestar$};
\node (meltdownss) [large,fails,below of=meltdownud] {Meltdown-SS $\medwhitestar$};
\node (meltdownbr) [large,group,below of=meltdownss] {Meltdown-BR};
\node (meltdowngp) [large,works,below of=meltdownbr] {Meltdown-GP~\cite{ARMSpecAnalysis,Intel2018white_paper}};

\node (meltdownpfus) [large,works,right of=meltdownnm,xshift=3.75cm] {Meltdown-US~\cite{Lipp2018meltdown}};
\node (meltdownpfp) [large,works,below of=meltdownpfus] {Meltdown-P~\cite{Vanbulck2018,Weisse2018foreshadowNG}};
\node (meltdownpfrw) [large,works,below of=meltdownpfp] {Meltdown-RW~\cite{Kiriansky2018speculative}};
\node (meltdownpfpk) [large,works,below of=meltdownpfrw] {Meltdown-PK $\medstar$};
\node (meltdownpfxd) [large,fails,below of=meltdownpfpk] {Meltdown-XD $\medwhitestar$};
\node (meltdownpfsm) [large,fails,below of=meltdownpfxd] {Meltdown-SM $\medwhitestar$};

\node (meltdownbrmpx) [large,works,below of=meltdownpfsm] {Meltdown-MPX~\cite{IntelMitigations}};
\node (meltdownbrbnd) [large,works,below of=meltdownbrmpx] {Meltdown-BND $\medstar$};

\draw (root.east) edge[thick,->,>=stealth,in=-90,out=0] node[above,midway,xshift=-1cm,yshift=.3cm]
    {\includegraphics[height=4ex]{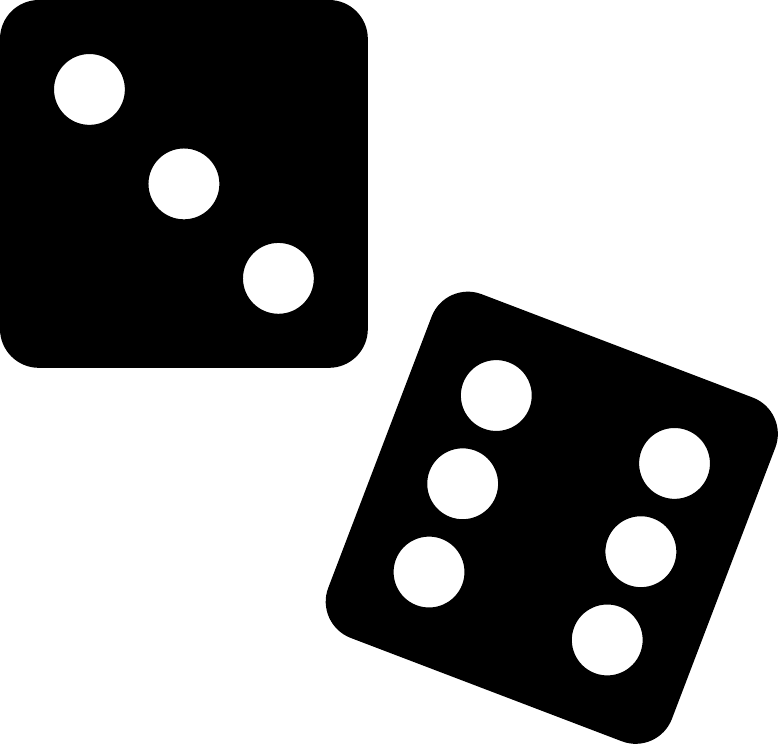}} node[decide, above,midway,xshift=-1cm,yshift=-.2cm] {prediction} (spectretype.south);
\draw (root.east) edge[thick,->,>=stealth,in=90,out=0] node[decide, below,midway,xshift=-1cm,yshift=.1cm] {fault} node[below,midway,xshift=-1cm,yshift=-.35cm]
    {\includegraphics[height=4ex]{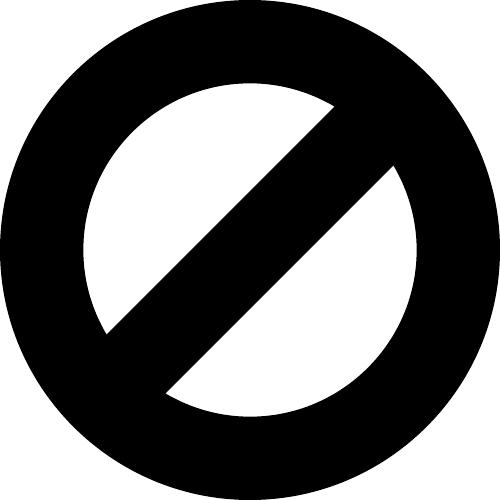}} (meltdowntype.north);

\draw (spectretype.east) edge[arrow_regular] (spectrepht.west);
\draw (spectretype.east) edge[arrow_regular] (spectrebtb.west);
\draw (spectretype.east) edge[arrow_regular] (spectrersb.west);
\draw (spectretype.east) edge[arrow_regular] (spectrestl.west);

\draw (spectrepht.east) edge[arrow_regular] (spectrephtca.west);
\draw (spectrepht.east) edge[arrow_regular] (spectrephtsa.west);
\draw (spectrephtca.east) edge[arrow_regular] (spectrephtcaip.west);
\draw (spectrephtca.east) edge[arrow_regular] (spectrephtcaop.west);
\draw (spectrephtsa.east) edge[arrow_regular] (spectrephtsaip.west);
\draw (spectrephtsa.east) edge[arrow_regular] (spectrephtsaop.west);

\draw (spectrebtb.east) edge[arrow_regular] (spectrebtbca.west);
\draw (spectrebtb.east) edge[arrow_regular] (spectrebtbsa.west);
\draw (spectrebtbca.east) edge[arrow_regular] (spectrebtbcaip.west);
\draw (spectrebtbca.east) edge[arrow_regular] (spectrebtbcaop.west);
\draw (spectrebtbsa.east) edge[arrow_regular] (spectrebtbsaip.west);
\draw (spectrebtbsa.east) edge[arrow_regular] (spectrebtbsaop.west);

\draw (spectrersb.east) edge[arrow] (spectrersbca.west);
\draw (spectrersb.east) edge[arrow] (spectrersbsa.west);
\draw (spectrersbca.east) edge[arrow_regular] (spectrersbcaip.west);
\draw (spectrersbca.east) edge[arrow_regular] (spectrersbcaop.west);
\draw (spectrersbsa.east) edge[arrow] (spectrersbsaip.west);
\draw (spectrersbsa.east) edge[arrow] (spectrersbsaop.west);

\draw (meltdowntype.east) edge[arrow_regular] (meltdownnm.west);
\draw (meltdowntype.east) edge[arrow_regular] (meltdownud.west);
\draw (meltdowntype.east) edge[arrow_regular] (meltdownde.west);
\draw (meltdowntype.east) edge[arrow_regular] (meltdownpf.west);
\draw (meltdowntype.east) edge[arrow_regular] (meltdownac.west);
\draw (meltdowntype.east) edge[arrow] (meltdownbr.west);
\draw (meltdowntype.east) edge[arrow] (meltdowngp.west);
\draw (meltdowntype.east) edge[arrow] (meltdownss.west);

\draw (meltdownpf.east) edge[arrow_regular] (meltdownpfus.west);
\draw (meltdownpf.east) edge[arrow_regular] (meltdownpfp.west);
\draw (meltdownpf.east) edge[arrow_regular] (meltdownpfrw.west);
\draw (meltdownpf.east) edge[arrow_regular] (meltdownpfpk.west);
\draw (meltdownpf.east) edge[arrow_regular] (meltdownpfxd.west);
\draw (meltdownpf.east) edge[arrow_regular] (meltdownpfsm.west);

\draw (meltdownbr.east) edge[arrow_regular] (meltdownbrmpx.west);
\draw (meltdownbr.east) edge[arrow_regular] (meltdownbrbnd.west);

\end{tikzpicture}

%% file: images/mistrain_location.tikz
\begin{tikzpicture}

\node[fill=green!20] at (-1.12,1.5) {\parbox{2cm}{\sl in-place/\\same-address-space}};
\node[fill=green!20] at (-1.12,3.75) {\parbox{2cm}{\sl out-of-place/\\same-address-space}};

\draw (0,0.75) rectangle +(3,3.75);
\node at (1.5,4.75) {\bf Victim};

\draw[fill=red!10] (0,1) rectangle +(3,1) node[pos=.5] {\parbox{3cm}{\centering Victim branch}};

\draw[fill=blue!10] (0,3.25) rectangle +(3,1) node[pos=.5] {\parbox{3cm}{\centering Congruent\\branch}};

\path[draw,>=stealth,-latex'] (1.5,3.25) to (1.5,2);
\path[draw,>=stealth,-latex'] (1.5,2) to node[rotate=90,midway,xshift=1.5em] {\parbox{2cm}{\scriptsize Address\\ collision}} (1.5,3.25);

\node[fill=green!20] at (8.1,1.5) {\parbox{2cm}{\sl in-place/\\cross-address-space}};
\node[fill=green!20] at (8.1,3.75) {\parbox{2cm}{\sl out-of-place/\\cross-address-space}};

\draw (4,0.75) rectangle +(3,3.75);
\node at (5.5,4.75) {\bf Attacker};

\draw[fill=yellow!10] (4,1) rectangle +(3,1) node[pos=.5] {\parbox{3cm}{\centering Shadow branch}};

\path[draw,>=stealth,-latex',densely dotted] (3,1.5) -> (4,1.5);
\path[draw,>=stealth,-latex',densely dotted] (4,1.5) -> (3,1.5);

\path[draw,>=stealth,-latex',densely dotted] (3,3.75) -> (4,3.75);
\path[draw,>=stealth,-latex',densely dotted] (4,3.75) -> (3,3.75);

\draw[fill=gray!10] (4,3.25) rectangle +(3,1) node[pos=.5] {\parbox{3cm}{\centering Congruent\\branch}};

\path[draw,>=stealth,-latex'] (5.5,3.25) to (5.5,2);
\path[draw,>=stealth,-latex'] (5.5,2) to node[rotate=90,midway,xshift=1.5em] {\parbox{2cm}{\scriptsize Address\\ collision}} (5.5,3.25);

\draw[fill=orange!20] (0,0) rectangle +(7,0.5) node[pos=.5] {Shared Branch Prediction State};

\end{tikzpicture}

%% file: ms.bbl
\begin{thebibliography}{10}

\bibitem{Aldaya2018}
{\sc Aldaya, A.~C., Brumley, B.~B., ul~Hassan, S., Garc\'ia, C.~P., and Tuveri,
  N.}
\newblock Port contention for fun and profit, 2018.

\bibitem{Aldaya2018PortSmash}
{\sc Aldaya, A.~C., Brumley, B.~B., ul~Hassan, S., Garc{\'{\i}}a, C.~P., and
  Tuveri, N.}
\newblock {Port Contention for Fun and Profit}.
\newblock {\em ePrint 2018/1060\/} (2018).

\bibitem{AMDssbd_whitepaper}
{\sc AMD}.
\newblock {AMD64 Technology: Speculative Store Bypass Disable}, 2018.
\newblock Revision 5.21.18.

\bibitem{AMDSpecAnalysis}
{\sc {AMD}}.
\newblock {Software Techniques for Managing Speculation on {AMD} Processors},
  2018.
\newblock Revison 7.10.18.

\bibitem{AMD2018_variant_3a}
{\sc AMD}.
\newblock Spectre mitigation update, July 2018.

\bibitem{ARMSpecAnalysis_whitepaper}
{\sc {ARM}}.
\newblock {Cache Speculation Side-channels}, 2018.
\newblock Version 2.4.

\bibitem{arm_arch_manualv85}
{\sc {ARM Limited}}.
\newblock {ARM A64 Instruction Set Architecture}, Sep 2018.

\bibitem{ARMSpecAnalysis}
{\sc {ARM Limited}}.
\newblock {Vulnerability of Speculative Processors to Cache Timing Side-Channel
  Mechanism}, 2018.

\bibitem{Bhattacharyya2019smother}
{\sc Bhattacharyya, A., Sandulescu, A., Neugschwandtner, M., Sorniotti, A.,
  Falsafi, B., Payer, M., and Kurmus, A.}
\newblock Smotherspectre: exploiting speculative execution through port
  contention.
\newblock {\em arXiv:1903.01843\/} (2019).

\bibitem{Carpenter2018smatch}
{\sc Carpenter, D.}
\newblock {Smatch check for Spectre stuff}, Apr. 2018.

\bibitem{Carruth2018retpoline_patch}
{\sc Carruth, C.}, \url{https://reviews.llvm.org/D41723}
Jan. 2018.

\bibitem{Carruth2018Hardening}
{\sc Carruth, C.}
\newblock {RFC: Speculative Load Hardening (a Spectre variant \#1 mitigation)},
  Mar. 2018.

\bibitem{Chen2018SGXpectre}
{\sc Chen, G., Chen, S., Xiao, Y., Zhang, Y., Lin, Z., and Lai, T.~H.}
\newblock {SGXPECTRE Attacks: Leaking Enclave Secrets via Speculative
  Execution}.
\newblock {\em arXiv:1802.09085\/} (2018).

\bibitem{Microsoft2019retpoline}
{\sc Corp., M.},
  \url{https://support.microsoft.com/en-us/help/4482887/windows-10-update-kb4482887}
Mar. 2019.

\bibitem{Culbertson2018performance}
{\sc Culbertson, L.}
\newblock Addressing new research for side-channel analysis.
\newblock Intel.

\bibitem{Dong2018Spectres}
{\sc Dong, X., Shen, Z., Criswell, J., Cox, A., and Dwarkadas, S.}
\newblock {Spectres, virtual ghosts, and hardware support}.
\newblock In {\em {Workshop on Hardware and Architectural Support for Security
  and Privacy}\/} (2018).

\bibitem{gcc2018speculation}
{\sc Earnshaw, R.}
\newblock {Mitigation against unsafe data speculation (CVE-2017-5753)}, July
  2018.

\bibitem{Evtyushkin2018BranchScope}
{\sc Evtyushkin, D., Riley, R., Abu-Ghazaleh, N.~C., ECE, and Ponomarev, D.}
\newblock Branchscope: A new side-channel attack on directional branch
  predictor.
\newblock In {\em ASPLOS'18\/} (2018).

\bibitem{Fog2016}
{\sc Fog, A.}
\newblock The microarchitecture of {Intel}, {AMD} and {VIA} {CPU}s: An
  optimization guide for assembly programmers and compiler makers, 2016.

\bibitem{Gregg2018kpti}
{\sc Gregg, B.}
\newblock {KPTI/KAISER Meltdown Initial Performance Regressions}, 2018.

\bibitem{Gruss2018Kernel}
{\sc Gruss, D., Hansen, D., and Gregg, B.}
\newblock Kernel isolation: From an academic idea to an efficient patch for
  every computer.
\newblock {\em USENIX ;login\/} (2018).

\bibitem{Gruss2017KASLR}
{\sc Gruss, D., Lipp, M., Schwarz, M., Fellner, R., Maurice, C., and Mangard,
  S.}
\newblock {KASLR is Dead: Long Live KASLR}.
\newblock In {\em {ESSoS}\/} (2017).

\bibitem{Gruss2016Prefetch}
{\sc Gruss, D., Maurice, C., Fogh, A., Lipp, M., and Mangard, S.}
\newblock {Prefetch Side-Channel Attacks: Bypassing SMAP and Kernel ASLR}.
\newblock In {\em CCS\/} (2016).

\bibitem{Gruss2015Template}
{\sc Gruss, D., Spreitzer, R., and Mangard, S.}
\newblock {Cache Template Attacks: Automating Attacks on Inclusive Last-Level
  Caches}.
\newblock In {\em USENIX Security Symposium\/} (2015).

\bibitem{Guarnieri2018spectector}
{\sc Guarnieri, M., K{\"{o}}pf, B., Morales, J.~F., Reineke, J., and
  S{\'{a}}nchez, A.}
\newblock {{SPECTECTOR:} Principled Detection of Speculative Information
  Flows}.
\newblock {\em arXiv:1812.08639\/} (2018).

\bibitem{Guelmezoglu2015}
{\sc G\"{u}lmezo\u{g}lu, B., Inci, M.~S., Eisenbarth, T., and Sunar, B.}
\newblock {A Faster and More Realistic Flush+Reload Attack on AES}.
\newblock In {\em {Constructive Side-Channel Analysis and Secure Design}\/}
  (2015).

\bibitem{Hedayati2018}
{\sc Hedayati, M., Gravani, S., Johnson, E., Criswell, J., Scott, M., Shen, K.,
  and Marty, M.}
\newblock {Janus: Intra-Process Isolation for High-Throughput Data Plane
  Libraries}, 2018.

\bibitem{Horn2018spectre}
{\sc Horn, J.}
\newblock {Reading privileged memory with a side-channel}, Jan. 2018.

\bibitem{Horn2018spectre4}
{\sc Horn, J.}
\newblock {speculative execution, variant 4: speculative store bypass}, 2018.

\bibitem{Intel_SGX}
{\sc {Intel}}.
\newblock {Intel Software Guard Extensions (Intel SGX)}, 2016.

\bibitem{Intel_vol3}
{\sc Intel}.
\newblock {Intel{$\textregistered$} 64 and IA-32 Architectures Software
  Developer{$\textquoteright$}s Manual, Volume 3 (3A, 3B \& 3C): System
  Programming Guide}.

\bibitem{Intel2018pku}
{\sc {Intel}}.
\newblock {Intel Xeon Processor Scalable Family Technical Overview}, Sept.
  2017.

\bibitem{Intel_opt}
{\sc Intel}.
\newblock {Intel{$\textregistered$} 64 and IA-32 Architectures Optimization
  Reference Manual}, 2017.

\bibitem{Intel2018l1tf}
{\sc {Intel}}.
\newblock {Deep Dive: Intel Analysis of L1 Terminal Fault}, Aug. 2018.

\bibitem{Intel2018white_paper}
{\sc {Intel}}.
\newblock {Intel Analysis of Speculative Execution Side Channels }, July 2018.
\newblock Revision 4.0.

\bibitem{Intel2018sok_mitigation}
{\sc {Intel}}.
\newblock {More Information on Transient Execution Findings},
  \url{https://software.intel.com/security-software-guidance/insights/more-information-transient-execution-findings}
2018.

\bibitem{Intel2018m3a}
{\sc {Intel}}.
\newblock {Q2 2018 Speculative Execution Side Channel Update}, May 2018.

\bibitem{Intel2018L1TFperf}
{\sc {Intel}}.
\newblock {Resources and Response to Side Channel L1 Terminal Fault}, Aug.
  2018.

\bibitem{Intel2018retpoline}
{\sc {Intel}}.
\newblock {Retpoline: A Branch Target Injection Mitigation}, June 2018.
\newblock Revision 003.

\bibitem{IntelMitigations}
{\sc {Intel}}.
\newblock {Speculative Execution Side Channel Mitigations}, May 2018.
\newblock Revision 3.0.

\bibitem{Ionescu2017TwitterApple}
{\sc Ionescu, A.}
\newblock {Twitter: Apple Double Map},
  \url{https://twitter.com/aionescu/status/948609809540046849}
2017.

\bibitem{Ionescu2017Twitter}
{\sc Ionescu, A.}
\newblock {Windows 17035 Kernel ASLR/VA Isolation In Practice (like Linux
  KAISER).}, \url{https://twitter.com/aionescu/status/930412525111296000}
2017.

\bibitem{Irazoqui2014}
{\sc Irazoqui, G., Inci, M.~S., Eisenbarth, T., and Sunar, B.}
\newblock {Wait a minute! A fast, Cross-VM attack on AES}.
\newblock In {\em RAID'14\/} (2014).

\bibitem{Islam2019Spoiler}
{\sc Islam, S., Moghimi, A., Bruhns, I., Krebbel, M., Gulmezoglu, B.,
  Eisenbarth, T., and Sunar, B.}
\newblock {SPOILER: Speculative Load Hazards Boost Rowhammer and Cache
  Attacks}.
\newblock {\em arXiv:1903.00446\/} (2019).

\bibitem{khasawneh2018safespec}
{\sc Khasawneh, K.~N., Koruyeh, E.~M., Song, C., Evtyushkin, D., Ponomarev, D.,
  and Abu-Ghazaleh, N.}
\newblock {SafeSpec: Banishing the Spectre of a Meltdown with Leakage-Free
  Speculation}.
\newblock {\em arXiv:1806.05179\/} (2018).

\bibitem{ARMSpectreLinuxv2Hardening}
{\sc King, R.}
\newblock {ARM: spectre-v2: harden branch predictor on context switches}, May
  2018.

\bibitem{Kiriansky2018dawg}
{\sc Kiriansky, V., Lebedev, I., Amarasinghe, S., Devadas, S., and Emer, J.}
\newblock {DAWG: A Defense Against Cache Timing Attacks in Speculative
  Execution Processors}.
\newblock {\em ePrint 2018/418\/} (May 2018).

\bibitem{Kiriansky2018speculative}
{\sc Kiriansky, V., and Waldspurger, C.}
\newblock {Speculative Buffer Overflows: Attacks and Defenses}.
\newblock {\em arXiv:1807.03757\/} (2018).

\bibitem{Kocher2018mitigations}
{\sc Kocher, P.}
\newblock Spectre mitigations in {Microsoft's} {C/C++} compiler, 2018.

\bibitem{Kocher2019spectre}
{\sc Kocher, P., Horn, J., Fogh, A., Genkin, D., Gruss, D., Haas, W., Hamburg,
  M., Lipp, M., Mangard, S., Prescher, T., Schwarz, M., and Yarom, Y.}
\newblock Spectre attacks: Exploiting speculative execution.
\newblock In {\em S\&P\/} (2019).

\bibitem{Kocher1996}
{\sc Kocher, P.~C.}
\newblock {Timing Attacks on Implementations of Diffe-Hellman, RSA, DSS, and
  Other Systems}.
\newblock In {\em CRYPTO\/} (1996).

\bibitem{Koruyeh2018spectre5}
{\sc Koruyeh, E.~M., Khasawneh, K., Song, C., and Abu-Ghazaleh, N.}
\newblock {Spectre Returns! Speculation Attacks using the Return Stack Buffer}.
\newblock In {\em {WOOT}\/} (2018).

\bibitem{Larabel2018stibp}
{\sc Larabel, M.}
\newblock {Bisected: The Unfortunate Reason Linux 4.20 Is Running Slower}, Nov.
  2018.

\bibitem{Larabel2018cost}
{\sc Larabel, M.}
\newblock The performance cost of spectre / meltdown / foreshadow mitigations
  on linux 4.19, Aug. 2018.

\bibitem{Lipp2016}
{\sc {Lipp}, M., {Gruss}, D., {Spreitzer}, R., Maurice, C., and {Mangard}, S.}
\newblock {ARMageddon: Cache Attacks on Mobile Devices}.
\newblock In {\em USENIX Security Symposium\/} (2016).

\bibitem{Lipp2018meltdown}
{\sc Lipp, M., Schwarz, M., Gruss, D., Prescher, T., Haas, W., Fogh, A., Horn,
  J., Mangard, S., Kocher, P., Genkin, D., Yarom, Y., and Hamburg, M.}
\newblock {Meltdown: Reading Kernel Memory from User Space}.
\newblock In {\em USENIX Security Symposium\/} (2018).

\bibitem{lklm18eagerswitching}
{\sc Lutomirski, A.}
\newblock {x86/fpu: Hard-disable lazy FPU mode}, June 2018.

\bibitem{LWN_kpti}
{\sc LWN}.
\newblock The current state of kernel page-table isolation,
  \url{https://lwn.net/SubscriberLink/741878/eb6c9d3913d7cb2b/}
Dec. 2017.

\bibitem{Maisuradze2018spectre5}
{\sc {Maisuradze}, G., and {Rossow}, C.}
\newblock ret2spec: Speculative execution using return stack buffers.
\newblock In {\em CCS\/} (2018).

\bibitem{Maurice2017Hello}
{\sc Maurice, C., Weber, M., Schwarz, M., Giner, L., Gruss, D., Alberto~Boano,
  C., Mangard, S., and R{\"o}mer, K.}
\newblock {Hello from the Other Side: SSH over Robust Cache Covert Channels in
  the Cloud}.
\newblock In {\em NDSS\/} (2017).

\bibitem{Microsoft2018edge}
{\sc {Microsoft}}.
\newblock {Mitigating speculative execution side-channel attacks in Microsoft
  Edge and Internet Explorer}, Jan. 2018.

\bibitem{Miller2018taxonomy}
{\sc Miller, M.}
\newblock {Mitigating speculative execution side channel hardware
  vulnerabilities}, Mar. 2018.

\bibitem{OKeeffe18sgxspectre}
{\sc O'Keeffe, D., Muthukumaran, D., Aublin, P.-L., Kelbert, F., Priebe, C.,
  Lind, J., Zhu, H., and Pietzuch, P.}
\newblock {Spectre attack against SGX enclave}, Jan. 2018.

\bibitem{Oleksenko2017mpx}
{\sc Oleksenko, O., Kuvaiskii, D., Bhatotia, P., Felber, P., and Fetzer, C.}
\newblock {Intel {MPX} Explained: An Empirical Study of Intel {MPX} and
  Software-based Bounds Checking Approaches}.
\newblock {\em arXiv:1702.00719\/} (2017).

\bibitem{Oleksenko2018ysnb}
{\sc Oleksenko, O., Trach, B., Reiher, T., Silberstein, M., and Fetzer, C.}
\newblock {You Shall Not Bypass: Employing data dependencies to prevent Bounds
  Check Bypass}.
\newblock {\em arXiv:1805.08506\/} (2018).

\bibitem{Grsecurity2018Respectre}
{\sc {Open Source Security Inc}}.
\newblock Respectre{$\texttrademark$}: The state of the art in spectre
  defenses, Oct. 2018.

\bibitem{Osvik2006}
{\sc Osvik, D.~A., Shamir, A., and Tromer, E.}
\newblock {Cache Attacks and Countermeasures: the Case of AES}.
\newblock In {\em CT-RSA\/} (2006).

\bibitem{Microsoft2018Spectre}
{\sc Pardoe, A.}
\newblock Spectre mitigations in {MSVC}, 2018.

\bibitem{Pessl2016}
{\sc Pessl, P., Gruss, D., Maurice, C., Schwarz, M., and Mangard, S.}
\newblock {DRAMA: Exploiting DRAM Addressing for Cross-CPU Attacks}.
\newblock In {\em USENIX Security Symposium\/} (2016).

\bibitem{Pizlo18}
{\sc Pizlo, F.}
\newblock What {Spectre} and {Meltdown} mean for {WebKit}, Jan. 2018.

\bibitem{Schwarz2018JSZ}
{\sc Schwarz, M., Lipp, M., and Gruss, D.}
\newblock {JavaScript Zero: Real JavaScript and Zero Side-Channel Attacks}.
\newblock In {\em NDSS\/} (2018).

\bibitem{Schwarz2018KeyDrown}
{\sc Schwarz, M., Lipp, M., Gruss, D., Weiser, S., Maurice, C., Spreitzer, R.,
  and Mangard, S.}
\newblock {KeyDrown: Eliminating Software-Based Keystroke Timing Side-Channel
  Attacks}.
\newblock In {\em NDSS\/} (2018).

\bibitem{Schwarz2017Timers}
{\sc Schwarz, M., Maurice, C., Gruss, D., and Mangard, S.}
\newblock {Fantastic Timers and Where to Find Them: High-Resolution
  Microarchitectural Attacks in JavaScript}.
\newblock In {\em FC\/} (2017).

\bibitem{Schwarz2018netspectre}
{\sc Schwarz, M., Schwarzl, M., Lipp, M., and Gruss, D.}
\newblock {NetSpectre: Read Arbitrary Memory over Network}.
\newblock {\em arXiv:1807.10535\/} (2018).

\bibitem{Shacham2007}
{\sc Shacham, H.}
\newblock The geometry of innocent flesh on the bone: Return-into-libc without
  function calls (on the x86).
\newblock In {\em CCS\/} (2007).

\bibitem{Shih2017tsgx}
{\sc Shih, M.-W., Lee, S., Kim, T., and Peinado, M.}
\newblock {T-SGX: Eradicating controlled-channel attacks against enclave
  programs}.
\newblock In {\em NDSS\/} (2017).

\bibitem{Chromium18enablebuffers}
{\sc Smith, B.}
\newblock {Enable SharedArrayBuffer by default on non-android}, Aug. 2018.

\bibitem{Stecklina2018LazyFP}
{\sc Stecklina, J., and Prescher, T.}
\newblock {LazyFP: Leaking {FPU} Register State using Microarchitectural
  Side-Channels}.
\newblock {\em arXiv:1806.07480\/} (2018).

\bibitem{Suse2018DisableBP}
{\sc {SUSE}}.
\newblock {Security update for kernel-firmware},
  \url{https://www.suse.com/support/update/announcement/2018/suse-su-20180008-1/}
2018.

\bibitem{Chromium18mitigations}
{\sc {The Chromium Projects}}.
\newblock {Actions required to mitigate Speculative Side-Channel Attack
  techniques}, 2018.

\bibitem{Chromium2018SiteIsolation}
{\sc {The Chromium Projects}}.
\newblock {Site Isolation}, 2018.

\bibitem{Tkachenko2018ibrs_performance}
{\sc Tkachenko, V.}
\newblock {20-30\% Performance Hit from the Spectre Bug Fix on Ubuntu}, Jan.
  2018.

\bibitem{Turner2018retpoline}
{\sc Turner, P.}
\newblock {Retpoline: a software construct for preventing
  branch-target-injection}, 2018.

\bibitem{VahldiekOberwagner2018}
{\sc Vahldiek{-}Oberwagner, A., Elnikety, E., Garg, D., and Druschel, P.}
\newblock {ERIM:} secure and efficient in-process isolation with memory
  protection keys.
\newblock {\em arXiv:1801.06822\/} (2018).

\bibitem{Vanbulck2018}
{\sc Van~Bulck, J., Minkin, M., Weisse, O., Genkin, D., Kasikci, B., Piessens,
  F., Silberstein, M., Wenisch, T.~F., Yarom, Y., and Strackx, R.}
\newblock {Foreshadow: Extracting the Keys to the Intel SGX Kingdom with
  Transient Out-of-Order Execution}.
\newblock In {\em USENIX Security Symposium\/} (2018).

\bibitem{Vanbulck2018nemesis}
{\sc Van~Bulck, J., Piessens, F., and Strackx, R.}
\newblock Nemesis: Studying microarchitectural timing leaks in rudimentary
  {CPU} interrupt logic.
\newblock In {\em CCS\/} (2018).

\bibitem{Varda2018cloudflare}
{\sc Varda, K.}
\newblock {WebAssembly’s post-MVP future},
  \url{https://news.ycombinator.com/item?id=18279791}
2018.

\bibitem{Wagner18firefox}
{\sc Wagner, L.}
\newblock Mitigations landing for new class of timing attack, Jan. 2018.

\bibitem{Wang2017oo7}
{\sc Wang, G., Chattopadhyay, S., Gotovchits, I., Mitra, T., and Roychoudhury,
  A.}
\newblock {{oo7:} Low-overhead Defense against Spectre Attacks via Binary
  Analysis}.
\newblock {\em arXiv:1807.05843\/} (2018).

\bibitem{Weisse2018foreshadowNG}
{\sc Weisse, O., Van~Bulck, J., Minkin, M., Genkin, D., Kasikci, B., Piessens,
  F., Silberstein, M., Strackx, R., Wenisch, T.~F., and Yarom, Y.}
\newblock {Foreshadow-NG: Breaking the Virtual Memory Abstraction with
  Transient Out-of-Order Execution}, 2018.

\bibitem{Yan2018InvisiSpec}
{\sc Yan, M., Choi, J., Skarlatos, D., Morrison, A., Fletcher, C.~W., and
  Torrellas, J.}
\newblock {InvisiSpec: Making Speculative Execution Invisible in the Cache
  Hierarchy}.
\newblock In {\em MICRO\/} (2018).

\bibitem{Yarom2014}
{\sc Yarom, Y., and Falkner, K.}
\newblock {Flush+Reload: a High Resolution, Low Noise, L3 Cache Side-Channel
  Attack}.
\newblock In {\em USENIX Security Symposium\/} (2014).

\end{thebibliography}
